\documentclass[AMA,STIX1COL]{WileyNJD-v2}
\usepackage{moreverb}
\usepackage{amsmath,amssymb}   
\usepackage{latexsym}
\usepackage{enumerate}
\usepackage{epsfig}
\usepackage{graphicx}
\usepackage{color}
\usepackage{float}
\usepackage{subfigure}
\usepackage{makeidx}
\usepackage{fancyhdr}
\usepackage{lastpage}
\usepackage{url}
\usepackage{wrapfig}
\usepackage{tikz}
\usepackage{xcolor}
\usepackage{graphicx}
\usepackage{array}

\newcommand\BibTeX{{\rmfamily B\kern-.05em \textsc{i\kern-.025em b}\kern-.08em
T\kern-.1667em\lower.7ex\hbox{E}\kern-.125emX}}

\articletype{Article Type}%

\received{<day> <Month>, <year>}
\revised{<day> <Month>, <year>}
\accepted{<day> <Month>, <year>}
\date{\today}

\begin{document}

\title{Generalized functional linear regression models with a mixture of complex functional and scalar covariates prone to measurement error}

\author[1]{Yuanyuan Luan*}

\author[1]{Roger S. Zoh}

\author[2]{Sneha Jadhav}

\author[3]{Lan Xue}

\author[1] {Carmen D. Tekwe*}

\authormark{Luan \textsc{et al}}

\address[1]{\orgdiv{Department of Epidemiology and Biostatistics, School of Public Health}, \orgname{Indiana University}, \orgaddress{\state{Indiana}, \country{United States}}}

\address[2]{\orgdiv{Department of Mathematics and Statistics}, \orgname{Wake Forest University}, \orgaddress{\state{North Carolina},\country{United States}}}

\address[3]{Department of Statistics, Oregon State University, Corvallis, OR 97331\\}


\corres{*Yuanyuan Luan or Carmen Tekwe, Bloomington IN 47405. \email{luany@iu.edu or ctekwe@iu.edu}}


\abstract[Abstract]{Current methods for correcting measurement error in covariates in generalized linear regression models apply to scalar covariates only.  No prior method exists to correct for measurement error in a mixture of functional and scalar covariates. We present simulation extrapolation (SIMEX) and regression calibration (RC) approaches to correct measurement errors in a mixture of functional and scalar covariates prone to classical measurement error in generalized functional linear regression. We conducted extensive simulations to assess the finite sample performance of our methods. Across a range of conditions, the SIMEX estimator generally had low bias, similar to that of the oracle estimator. The RC estimator performed almost as well, and notably better than average and naive estimators that did not correct for measurement error. We applied our methods to data from the 2011-2014 cycles of the National Health and Examination Survey (NHANES) to assess the relationships physical activity, total caloric intake, and demographic characteristics have with type 2 diabetes in community-dwelling adults living in the United States. We treated the device-based measure of physical activity as a functional covariate prone to complex arbitrary heteroscedastic errors and total caloric intake as a scalar covariate prone to error. Estimation with the SIMEX and RC estimators reduced the associations for some covariates, but increased them for others, particularly physical activity, in comparison to estimation with the estimators that did not correct for measurement error. Our approaches to correcting measurement error may be useful for generalized functional linear regression in other empirical settings as well.} 



\maketitle


\section{Introduction}\label{sec1}

Generalized linear regression is a parametric approach to modeling associations between one or more scalar predictors and a scalar outcome under the exponential family framework\cite{nelder1972generalized}. Researchers have applied generalized linear regression models in various fields including biostatistics and epidemiology\cite{liu2019application,vergotine2014proliferator,joshi2021predicting}, where the outcomes of interest may appear as count, continuous, or binary variables. While generalized linear regression models focus on the association between scalar covariates and outcomes, statisticians have extended the models to include functional covariates\cite{james2002generalized,crainiceanu2009generalized,fan2014generalized}. Crainiceanu and colleagues\cite{crainiceanu2009generalized}  extended generalized multi-level mixed models to generalized multi-level functional linear models by including a multilevel functional predictor. Fan and colleagues\cite{fan2014generalized} used generalized functional linear regression to investigate the association between a dichotomous disease trait with multiple genetic variants in a genetic region. Gertheiss and colleagues\cite{gertheiss2013variable} applied the generalized functional linear regression model to assess the association between brain magnetic resonance imaging indices and multiple sclerosis, treating the indices along the same white matter tracts as correlated functional variables\cite{staicu2012modeling}.

Although functional data analysis is well-developed\cite{ramsay1991some,cardot2007smoothing, cardot2003spline, goldsmith2011penalized, crainiceanu2009generalized,yao2005functional,reiss2007functional,james2009functional,law2015analysis}, methods for the analysis of functional data contaminated with measurement error are an active area of research. In practice, researchers observe functional data discretely and assume they reflect underlying latent smooth functions\cite{ramsay2007applied}. Functional data analysts have assumed that functional covariates are directly observed, are measured without error, or have error in the form of white noise \cite{amini2012sampled,araki2009functional,boente2000kernel,cai2006prediction,cardot2005estimation}. Thus, researchers did not distinguish between errors as independent random noisy fluctuations around a smooth trajectory and classical measurement errors with complex structures\cite{wang2016functional, cardot2007smoothing, yao2005functional}. To address the noisy fluctuations, researchers have smoothed the functional covariates before including them in a model\cite{ramsay2005functional}. In many circumstances, the true functional covariates may not be directly observable or the measurements may be prone to measurement error\cite{tekwe2019instrumental}. Earlier treatments of measurement error in functional data analysis involved assuming the errors are discrete and uncorrelated along the smooth trajectory\cite{yao2005functional,rice2001nonparametric}. For example, Yao and colleagues\cite{yao2005functional} established a functional principal components analysis (FPCA) for addressing measurement errors in longitudinal functional data by assuming uncorrelated measurement error covariance structures. Cardot and colleagues\cite{cardot2007smoothing} corrected for measurement error in functional data but considered the error as white noise with a constant variance. We also developed FPCA-based methods for reducing biases in a functional covariate prone to error for the measurement error exposure model for the functional multiple indicators, multiple causes measurement error models\cite{tekwe2018functional}. For these methods, we also assumed an independent error structure. Although these prior methods correct for some types of measurement error in functional covariates, they do not address complex measurement error structures, such as serially observed functional data with errors correlated within subjects and between observation periods. 

Other researchers have proposed approaches to overcome these limitations. Crainiceanu et al. \cite{crainiceanu2009generalized} demonstrated that the generalized multilevel functional linear models may be analyzed in a generalized multilevel mixed models framework. The authors consider the models the functional analog of measurement error models and used multilevel FPCA for the measurement error component of the model. Chakraborty and Panaretos\cite{chakraborty2017regression} described regression calibration-based methods for correcting measurement error that account for serial correlations in error structures. They assumed serial correlations exist within a small grid interval or a banded covariance structure for measurement errors. Thus, they assumed that the correlations depended on the distance between the observations. More recently, we presented an approach to correcting for measurement error in functional linear regression, demonstrating that "pre-smoothing" approaches do not correct for measurement errors with complex serial correlation structures adequately\cite{tekwe2019instrumental} . For this method and another method\cite{jadhav2022function}, we used functional instrumental variables belonging in the same time space as the true covariate to identify the model and correct for measurement error. In addition, we developed a Bayesian approach to correcting measurement error in functional linear regression in which we assumed a truncated Dirichlet Process mixture prior for the measurement error and also used an instrumental variable\cite{zoh2022fully}. However, practical shortcomings of using an instrumental variable are the required assumption that it belongs in the same time space as the observed functional covariate, which may not always be feasible\cite{tekwe2019instrumental}, and the frequent difficulty of identifying a good instrument. Cai\cite{caimethods} proposed several methods for correcting measurement error in functional covariates in scalar-on-function regression under different assumptions about the measurement error covariance matrix. These assumptions include constant variance, heteroscedastic subject-specific variance, and an auto-correlated covariance structure that depends on a set of parameters. For the constant variance and subject-specific variance assumptions, measurement errors are not correlated across time points, a condition that functional data prone to measurement error do not meet\cite{tekwe2019instrumental}.

Existing methods for correcting measurement error in functional regression focus on measurement error in a single function-valued covariate \cite{crainiceanu2009generalized,jadhav2020functional,chakraborty2017regression,zoh2022fully}. Sometimes, however, researchers seek to assess how mixtures of error-prone functional and scalar covariates influence either discrete or continuous outcomes. For example, in our current work, we evaluate the relationships physical activity and total caloric intake have with type 2 diabetes (T2D). Wearable accelerometer-based devices are now the standard tools for  monitoring and quantifying physical activity objectively. However, these devices also generate data prone to error \cite{case2015accuracy,freedson2005calibration, corder2007accelerometers}. For instance, step counts recorded by wearable devices (Nike Fuelband) can differ by up to $22\%$ from actual step counts \cite{case2015accuracy}. 
Self-reports of dietary intake are the most common and practical method for assessing caloric intake, especially in large surveys. However, measurement error in such reports stemming from social desirability, forgetting, and other factors\cite{bailey2021overview,subar2015addressing,dwyer1989memory} obscures the relationship between diet and health\cite{archer2013validity, mitka2013flawed, archer2015inadmissibility, adusumilli2020inference}. Moreover, when the calculation of total caloric intake is based on many separate self-reports of particular foods and beverages consumed, the measurement errors accumulate \cite{subar2015addressing}. 

We present a new method that involves parametric and semiparametric approaches to correct for measurement error in functional and scalar covariates in functional linear regression. In Section 2, we define the generalized functional linear regression model with mixed measurement error. In Section 3, we describe our proposed measurement error correction methods. In Section 4, we report on the performance of our methods in extensive simulations. In Section 5, we describe results from applying our method in the analysis of data on physical activity, caloric intake, and T2D from the National Health and Nutrition Examination Survey. In Section 6, we conclude with a discussion of our method and results.

\section{Generalized functional linear regression model with mixed measurement error}
\subsection{The model}

In the following generalized functional linear regression model with mixed measurement error (GFLM-ME), we describe the relation between $Y \in \mathbb{R}$ and the functional variable $X_1(\cdot)\in L^2[0,1]$ and the scalar variables $X_2 \in \mathbb{R}$ and $\boldsymbol{Z} \in \mathbb{R}$. Suppose  $\left\{y_i\right\}^{n}_{i=1}$ is a realization of random variables $\left\{Y_i\right\}^{n}_{i=1} \in \mathbb{R}$, which are independently distributed from an exponential family with mean $\mu_i$. In the GFLM-ME, the conditional mean response $\mu_{i}=E\{y_i|X_{1i}(t),X_{2i},{\bf Z}_{i}\}$ for the $i$th ($i=1, \hdots, n$) subject is  
\begin{align}
g(\mu_{i}) &=\int_{0}^{1} \beta_1(t) X_{1i}(t) dt + \beta_2 X_{2i}  +\boldsymbol{\alpha} \mathbf{Z}_{i},  \label{eq:1} \\
W_{1i}(t) &=X_{1i}(t)+U_{1i}(t), \text{ and} \label{eq:2} \\
W_{2i} &= X_{2i} + U_{2i}, \label{eq:3}
\end{align}
where $\beta_1(t) \in L^2[0,1]$ is an unknown functional coefficient and $\beta_2$ is an unknown scalar coefficient. The $\boldsymbol{\alpha}$ represents a $1 \times p$ vector of unknown scalar coefficients for the error-free covariates and $g(\cdot)$ represents the link function that is monotone and twice continuously differentiable with bounded derivatives \cite{nelder1972generalized, muller2005generalized}. The outcome, $Y_i$, is a continuous or discrete scalar random variable, and $X_{1i}(t), t\in [0,1]$ is a random function and assumed to be integrable on the unit time interval $[0,1]$. Additionally, $X_{1i}(t)$ and $X_{2i}$ are not directly measured or observed but are approximated by surrogate measures, $W_{1i}(t)$ and $W_{2i}$, respectively. $W_{1i}(t)$ and $W_{2i}$ are prone to measurement errors, $U_{1i}(t)$ and $U_{2i}$, respectively.  The $\boldsymbol{Z}_i \in \mathbb{R}^p$ is a $p \times 1$ vector that contains $p$ error-free scalar covariates. The model consists of three components: a response model for $Y_i$ in  (\ref{eq:1}) and measurement error models for $W_{1i}(t)$  and $W_{2i}$ in (\ref{eq:2}) and (\ref{eq:3}), respectively.

\subsection{Estimation} \label{estimate}

The presence of measurement error in regression models makes them unidentifiable without additional data or identifying information \cite{fuller2009measurement,carroll2006measurement}. Because the true covariates $X_{1i}(t)$ and $X_{2i}$ are not observed, and the corresponding measurement error terms in (\ref{eq:2}) and (\ref{eq:3}) are unknown, estimation of the covariance matrices associated with the measurement errors requires additional data. The additional data may appear as instrumental variables or repeated measures on the surrogate data for the true covariates. The measurement error covariance matrix can be assumed to be known if it was previously estimated from another data set or from additional identifying or validation data on a subset of subjects.

We use repeated measures to estimate the measurement error covariance matrix. In this approach, the model for the $i${th} subject at the $j_1${th} $j_1 = (1,\cdots,J_1)$ and $j_2${th} $j_2 = (1,\cdots,J_2)$ repeated session for $W_{1ij_1}(t)$ and $W_{2ij_2}$ in (\ref{eq:2}) and  (\ref{eq:3}), respectively, becomes
\begin{align}
W_{1ij_1}(t) &=X_{1i}(t)+U_{1ij}(t) \quad \quad j_1 =1, \ldots, J_1, \text{ and} \label{eq:4} \\
W_{2ij_2} &= X_{2i} + U_{2ij_2} \quad \quad j_2 =1, \ldots, J_2, \label{eq:5}
\end{align}
where $J_1$ and $J_2$ are the number of repeated measures for $W_{i1}(t)$ and $W_{i2}$, respectively. In practice, the number of repeated measures $J_1$ and $J_2$ can be different. For notational simplicity, we assume $J_1=J_2=J$. 

In (\ref{eq:1}), $\beta_1(t)$ is infinite-dimensional, which presents a challenge for estimation. Therefore, we first approximate $\beta_1(t)$ by reducing its dimension through polynomial spline basis expansion. The polynomial spline is flexible, captures the main features of the curve, and has desirable theoretical properties, allowing the approximation of a functional variable with a finite number of known basis functions \cite{ramsay2004functional}. The spline basis expansion gives $\beta_1(t) \approx \sum_{k=1}^{K_n} \gamma_k b_k(t)$, where $\left\{\gamma_k\right\}_{k=1}^{K_n}$ are unknown spline coefficients and $\left\{b_k(t)\right\}_{k=1}^{K_n}$ is a set of spline basis functions defined on the unit interval. The number of bases, $K_n$, is selected based on the sample size $n$ to provide a reasonable approximation of $\beta_1(t)$.

The proposed models in (\ref{eq:1}), (\ref{eq:4}), and (\ref{eq:5}) become
\begin{align}
g \left(\mu_{i}\right) &\approx \sum_{k=1}^{K_n} \gamma_k {X_{1ik}} + \beta_2 X_{2i}  +\boldsymbol{\alpha} \mathbf{Z}_{i}  \label{eq:6}, \\
W_{1ijk} &=X_{1ik} +U_{1ijk} \quad \quad k = 1, \dots , K_n \label{eq:7}, \text{ and} \\
W_{2ij} &= X_{2i} + U_{2ij}, \label{eq:8}
\end{align}
where  $W_{1ijk}=\int_{0}^{1} W_{1ij}(t) b_{k}(t) dt$, $X_{1ik}=\int_{0}^{1} X_{1i}(t) b_{k}(t) dt$, and  $U_{1ijk}=\int_{0}^{1} U_{1ij}(t) b_{k}(t) dt$. The reduced model is a generalized linear regression model in which the number of coefficients depends on sample size. 

\subsection{Assumptions}
We make the following assumptions for our methods. 
\begin{enumerate}
\item The observed covariate is an unbiased surrogate for the true covariate. That is, $E\{W_{1ij}(t)|X_{1i}(t)\} = X_{1i}(t),E(W_{2ij}|X_{2i}) = X_{2i}, i=1,\dots, n, \text{and } j =1, \dots, J$.

\item The true covariates and measurement errors are uncorrelated. That is, 
	$Cov\{X_{1i}(t), U_{1ij}(t)\} = 0 \text{ and }  Cov\{X_{2i}, U_{2ij}\} = 0$, for $t\in [0,1], i=1,\dots, n,$ \text{and} $ j =1 \dots, J$. 
 
\item The observed covariates and measurement errors are correlated, that is, $Cov\{W_{1ij}(t), U_{1ij}(t)\} \ne 0, i=1,\dots, n, \text{ and } j =1, \dots, J$ for any $t\in [0,1]$, and $Cov\{W_{2ij}, U_{2ij}\} \ne 0, i=1,\dots, n, \text{ and } j =1, \dots, J$. 

\item The measurement errors $U_{1ij}(t
)$ and $U_{2ij}$ are uncorrelated for any $t \in [0,1]$, $i=1,\dots,n, \text{ and } j =1, \dots, J$.

\item The functional measurement error, $U_{1ij}(t)$, is a Gaussian process with zero mean function and covariance function denoted as $ Cov\{U_1(t), U_1(t^{\prime})\}= \Sigma_{U_1}(t,t^{\prime})$ for $t \ne t^{\prime}$.
\item The measurement error for the scalar covariate follows the Gaussian distribution, i.e., $U_{2ij}\overset{\text{iid}}{\sim} N ({0}, {\sigma}^2_{U_{2}}), i=1,\dots, n, \text{ and } j =1, \dots, J$.

\end{enumerate}

Although assumptions 1-4 are necessary in classical additive measurement error models \cite{carroll2006measurement, buonaccorsi2010measurement}, verifying them in practice can be challenging because measurement errors are not directly observable. However, assumptions 5 and 6 about the normality of measurement errors $U_{1ij}(t)$ and $U_{2ij}$ can be relaxed to allow non-symmetric distributions. 

\section{Method and Estimation}

\subsection{Simulation Extrapolation}

Simulation extrapolation (SIMEX), proposed by Cook and Stefanski \cite{cook1994simulation}, is a method for estimating the effect of a covariate contaminated with measurement error in regression models. While originally developed for scalar variables, SIMEX can also be applied to functional variables \cite{caimethods,tekwe2022estimation}. An advantage of SIMEX is that it does not require an assumed or known distribution of the true covariate \cite{carroll2006measurement}. However, it does require the variance of the measurement error to be known or estimated from additional data \cite{carroll2006measurement, cook1994simulation}. Cook and Stefanski \cite{cook1994simulation} assumed homoscedastic and known measurement error variance. We assume that the measurement error covariance in the functional covariate follows a Gaussian process as in assumption 5, and estimate the measurement error variance using repeated measures of the covariates that are prone to measurement error. We also assume that $U_{1i_1j_1}(t)$ is independent of $U_{1i_2j_2}(t)$ where $i_1 \neq i_2, \text{and} j_1 \neq j_2$

SIMEX is easy to understand and implement, involving two main stages: simulation and extrapolation. In the simulation stage, additional measurement errors -- pseudo errors $U^*$ -- are generated from an assumed distribution and added to the originally observed measures, $W$, with a positive tuning parameter $\lambda$, resulting in new observed measures $W^*$ with successively larger measurement errors, where $W^* = W+ \lambda U^*$. The $\lambda$ controls the size of the measurement error added to $W$. The estimated regression parameters are then obtained from $W^*$. These simulation steps are repeated a large number of times (e.g., 200 times), yielding average values of the estimates for each value of $\lambda$. The extrapolation stage involves extrapolating the average estimates from the simulation stage to the case of no measurement error ($\lambda = -1$). The unbiasedness of the SIMEX estimator relies heavily on the accurate estimation of the variance of the measurement error and accurate assumption about the distribution of the measurement error. \cite{stefanski1995simulation}. 

We use the SIMEX method to estimate $\beta_1(t)$ and $\beta_2$ and $\boldsymbol{\alpha}$ in (\ref{eq:1}) for the generalized linear functional regression model with two covariates prone to measurement error. When the true measurements $X_{1i}(t)$ and $X_{2i}$ are both known, the corresponding estimated regression coefficients, $\widehat{\beta_{1}}(t)$ and $\widehat{{\beta_2}}$, are respectively consistent for the parameters in (\ref{eq:1}). Suppose $\boldsymbol{W}_{1i}(t)$ and $\boldsymbol{W}_{2i}$ are the observed measurements for the true covariates for subject $i$ across all the repeated sessions. Averaging the repeatedly observed $\boldsymbol{W}_{1i}(t)$ and $\boldsymbol{W}_{2i}$, across the replicates and using the resulting $\overline{{W}}_{1i}(t)$ and $\overline{{W_{2i}}}$ as substitutes for $X_{1i}(t)$ and $X_{2i}$, respectively, produces biased estimates of $\beta_1(t)$ and $\beta_2$, respectively.  

In our simulations and data application, we set $\lambda$ to be between 0 and 2, and express the pseudo-data as 
\begin{align*}
W_{1i, \lambda}(t) &= \overline{W}_{1i}(t) + \sqrt{\lambda}\overline{U}_{1i}(t), \text{ and}\\
W_{2i, \lambda} &= \overline{W}_{2i} + \sqrt{\lambda}\overline{U}_{2i},
\end{align*}
where the measurement errors, $\overline{U}_{1i}(t)$ and $\overline{U}_{2i}$, are averaged across replicates and simulated independently of $\{Y_i, \overline{W}_{1i}(t), \overline{W}_{2i}, \boldsymbol{Z}_{i}\}$ by Monte Carlo simulation and assumed to follow Gaussian process $\{\boldsymbol{0}, \widehat{\Sigma}_{U_1}(t,t^{\prime})\}$ for $t \ne t^{\prime}$ and $N(0, \widehat{\sigma}^2_{U_2})$, respectively. We use the replicates to estimate $\widehat{\Sigma}_{U_1}(t,t^{\prime})$ and $\widehat{\sigma}^2_{U_2}$ with
\begin{eqnarray*}
\widehat{\Sigma}_{U1}(t,t^{\prime}) &=& \frac{\sum_{i=1}^n \sum_{j=1}^J\left[W_{ij}(t^{\prime})-\bar{W}_i(t^{\prime})\right]\left[W_{ij}(t)-\bar{W}_i(t)\right]}{n (J-1)J} \text{ and}  \\
    \widehat{\sigma}_{u2}^{2} &=& \frac{\sum_{i=1}^n \sum_{j=1}^J\left[W_{ij}-\bar{W}_i\right]^2}{n (J-1)J}
\end{eqnarray*}
prior to the simulation stage. With the generated pseudo-data, we proceed with polynomial basis expansion as described in Subsection \ref{estimate} and model fitting to obtain $\widehat{\boldsymbol{\beta}}_{1\lambda}(t)$, $\widehat{{\beta_2}}_{\lambda}$, and $\widehat{\boldsymbol{\alpha}}_{\lambda}$, which are functions of $\lambda$. The simulation stage involves repeating the simulation of the pseudo-data and regression steps a total of $S$ times. We then average the estimated coefficients across all $S$ iterations for each $\lambda$ level\cite{carroll2006measurement}.

The extrapolation stage involves obtaining extrapolation equations for the averaged estimated coefficients as a function of $\lambda$. Thus, after averaging the estimated coefficients at each $\lambda$ across all $S$ iterations, we regress the average coefficients on functions of $\lambda$. The function of $\lambda$ may be linear, quadratic, or otherwise nonlinear. After fitting the regression models, we obtain the SIMEX-based estimator at $\lambda = -1$ and the naive estimator at $\lambda = 0$. In the Supplementary Materials, we summarize the steps for quadratic extrapolation when done for each covariate separately.

\subsection{Regression Calibration}
Carroll and Stefanski \cite{carroll1990approximate} and Gleser \cite{gleser1990improvements} proposed regression calibration (RC), which replaces the unobserved true measurement in the regression model with an unbiased approximation of it\cite{carroll2006measurement}. In practice, the unbiased approximation of the true measurement error can be obtained by various methods using internal validation, external validation, and replicate data. For example, a simple way to obtain the unbiased approximation of the true measurement error is to regress the true measurement, X,  on other covariates including error-free covariates, Z, and observed measurement of X, W, in the validation data. While regression calibration can be applied to any regression model, the method is particularly useful for generalized linear models \cite{carroll2006measurement}. There are several ways to obtain estimates of the true measurement under the regression calibration approach. When internal validation data are not available for all subjects in a sample, true measurement can be approximated with internal validation data from a subset of subjects in the sample data \cite{lee1995estimation}. Spiegelman and colleagues \cite{spiegelman1997regression} regressed the true measurement, $X$, on the observed measurement, $W$, using an internal validation data set to obtain an approximation of the true measurement on a subsample of individuals. With the measurement error covariance matrix estimated from external data or internal replicate data, the best linear approximation to $X$ can be obtained with a method-of-moments estimator \cite{carroll1990approximate,gleser1990improvements,hardin2003regression,liu1992efficacy,carroll1993case}. The James-Stein shrinkage estimator is another approach to approximating $X$ with replicate data on $W$ \cite{carroll2006measurement, whittemore1989errors}. It is close to the method-of-moments estimate of $X$ and involves estimating a reliability parameter, $R$, of the observation $W$, where $R=\frac{Var(X)}{Var(W)}$\cite{carroll2006measurement, BurghgraeveRC}, and shrinking $E(X|W)$, from $W$ towards $E(X)$ using $E(X|W) = RW+ (1-R) E(X)$\cite{BurghgraeveRC}. Expanded regression calibration models can also produce approximate measures of $X$. With identifying information to estimate the variance of the measurement error, expanded regression calibration involves simulating approximate measures for $X$ from its distribution, $f_{x}(X)$, or from its conditional distribution given observed data, namely from $f_{X|W,Z}(X|W,Z)$ \cite{carroll2006measurement}. This is especially useful when validation data are not available on a subset of subjects. For example, we \cite{tekwe2014multiple, tekwe2016generalized} developed Monte Carlo expectation maximization methods to simulate the true measurements from their conditional distribution given the observed data while using an instrumental variable to estimate the variance of the measurement error. 

We use here a variation of the expanded regression calibration method under the assumption that $E\{W_{ij}(t)\mid X_i(t)\} = X_i(t)$ and $cov[E\{W_{ij}(t)\mid X_i(t)\}] = \Sigma_{U_1}$ and consequently, $W_{ij}(t)\mid X_i(t) \sim MVN\{X_i(t), \Sigma_{U_1} \}$. Rather than simulating directly from the conditional distribution of the surrogate given the true covariate, Strand and colleagues \cite{strand2014regression} proposed a mixed effects approach to obtain recalibrated measures for $X_i(t)$ under the regression calibration method. This approach involves regressing the observed surrogates on their mean functions in a random intercept model. Given the dimension of longitudinal functional data, fitting a mixed effects model can be computationally intensive. To address these computational challenges, Cui and colleagues\cite{cui2021fast} proposed fast univariate inference (FUI), a scalable method for fitting massive longitudinal functional data. FUI requires fitting mixed-effect models on the observed covariate at each time point $t$, across all the repeated measures of $W$ over multiple days or periods of observation. Performing the mixed effects model in a univariate fashion creates a more scalable and less computationally intensive method for estimation. We use FUI to approximate true measures and use the repeated measures of the surrogate or observed data as identifying information. Under (\ref{eq:4}), for any fixed $t$,  
\begin{equation*}
W_{1ij}(t) =X_{1i}(t)+U_{1ij}(t)=\mu_{{X}_{1}}(t)+\epsilon_{1i}(t)+U_{1ij}(t), \quad \quad j =1, \cdots, J, i=1,\cdots, n,    
\end{equation*}
where $\mu_{{X}_{1}}(t)=E\{X_{1i}(t)\}$, and $\epsilon_{1i}(t)=X_{1i}(t)-\mu_{{X}_{1}}(t)$ is a mean zero random variable shared across $J$ repeated measures. It becomes a mixed-effects model with one subject-specific random intercept at each time point.
Therefore, FUI for regression calibration fits univariate mixed-effects models with $W_{1ij}(t)$ as outcomes at each time point to obtain a point-wise estimate of $X_{1i}(t)$, which may be subsequently smoothed along the functional domain. For the scalar covariate, a mixed-effects model can be used similarly to obtain estimates of  $X_{2i}$.  We implement this approach in several steps. First, we use FUI to estimate $X_{i1}(t)$ and $X_{i2}$, denoted by $\widehat{X}_{1i}(t)$ and $\widehat{X}_{2i}$, respectively. Next, we replace $X_{i1}(t)$ and $X_{i2}$ with $\widehat{X}_{1i}(t)$ and $\widehat{X}_{2i}$, respectively, in (\ref{eq:1}) to estimate coefficients corrected for measurement error. We obtain the RC-based estimators as follows.  

\begin{enumerate}
\item Obtain $\widehat{X}_{i1}(t)$ and $\widehat{X}_{i2}$ using FUI with separate linear mixed-effect regressions at each time point.  
    
\item Reduce the dimension of $\widehat{X}_{i1}(t)$ with basis expansion, converting $\widehat{X}_{i1}(t)$ into a $1\times K_n$ matrix. 

\item Perform generalized linear regression of ${Y}_{i}$ on $\widehat{X}_{i1}(t)$,  $\widehat{X}_{i2}$, and $\boldsymbol{Z}_{i}$ to obtain RC-based estimators for the coefficients, $\boldsymbol{\gamma}_{RC}$, $\beta_{2RC}$, and $\boldsymbol{\alpha}_{RC}$ as in (\ref{eq:6}). In particular, estimate the functional regression coefficient by 
\begin{eqnarray*}
\widehat{\beta}_{RC} (t) &\approx&  \boldsymbol{b}(t) \widehat{\boldsymbol{\gamma}}_{RC},
\end{eqnarray*}
where $\boldsymbol{b}(t)$ is a set of B-spline basis functions. 
\end{enumerate}

\subsection{Inference}
We used non-parametric bootstrap confidence intervals (CI) for inference. We first randomly selected subjects from the sample with replacement and then treated their data as the bootstrap sample. We estimated the parameters of interest under the different estimation methods using each bootstrap sample (see Section \ref{sec:sim}). We repeated these steps 500 times, resulting in 500 estimators for each method. The $95\%$ CIs for the functional coefficient are the $2.5$th and $97.5$th percentiles of $\widehat{\beta}_b(t)$ at each observed time point. We obtained the $95\%$ CIs for the scalar and error-free covariates as the $95\%$ percentile bootstrap CIs similarly.

\section{Simulations} \label{sec:sim}

\subsection{Simulation procedure}

We conducted simulations to investigate the finite sample performance of our method. We used the generalized functional linear regression model in (\ref{eq:1}) to generate binary responses. Specifically, we simulated the outcomes with $Y_i \sim Binomial(p_i)$ where $p_i = expit(\eta_i)$ and $\eta_i = \int_{0}^{1} \beta_1(t) X_{1i}(t) dt + \beta_2 X_{2i}  +\alpha_1 Z_{1i}+ \alpha_2 Z_{2i}$, in which $\beta_1(t) = sin(2\pi t)$, $\beta_2 =1$, and $\alpha_1=\alpha_2= 1$.
We simulated the true functional and scalar covariates with $X_{1i}(t)=1/[1+\exp\{8(t-0.5)\}] + \epsilon_{X_1}(t)$ for $t\in [0,1]$ and $X_{2i} \sim N(2, \sigma^2_{X_2})$, respectively.
 We simulated the error term, $\epsilon_{X_1}(t)$, independently from a mean zero Gaussian process with constant variances $\sigma^2_{X_1}$ and covariance function $\Sigma_{X_1}(s,t)=\sigma^{2}_{X_1}\exp\left\{-(s-t)^2/2l_{X}^2\right\}$, where the correlation depends on the distance between two points and $l_{X}$ controls the strength of correlation, with a smaller value of $l_{X}$ indicating a weaker correlation. In addition, we simulated the continuous and binary error-free covariates independently with $Z_{1i} \sim N(2, \sigma^2_Z)$ and $Z_{2i} \sim Binomial(1, 0.6)$, respectively. We simulated the observed functional covariate $W_{1ij}(t)$ from the additive measurement error model $W_{1ij}(t) = X_{1i}(t) + U_{1ij}(t)$, generating $U_{1ij}(t)$ from a mean zero Gaussian process with an exponential covariance function $\Sigma_{U_1}(s,t)=\sigma^{2}_{U_1}\exp\{-(s-t)^2/2l_{U}^2\}$, where  $l_{U}$ controls the strength of  correlation for the functional measurement error and $\sigma_{U_1}$ is the constant standard deviation of $U_1(t)$ along the functional span. Similarly, we simulated the observed scalar covariate $W_{2ij}$ as $W_{2ij}= X_{2i} + U_{2ij}$ and independently of $W_{1ij}(t)$, where $U_{2ij} {\sim} N(0, \sigma^2_{U_2})$.  Although the number of replicates for functional and scalar covariates prone to error can be different in practice, we used 5 replicates for both types of covariates in our simulation. 
 
We conducted simulations to evaluate the performance of the estimators under the following four sets of conditions: 1) increasing sample sizes of  $n \in (500, 1000, 2000, 5000)$ under different magnitudes of measurement error, $\sigma_U \in (1.5, 3)$; 2) different magnitudes of measurement error, $\sigma_{U} \in (1.5, 2, 3, 4)$ and under different correlations of functional measurement error, $l_U \in (0.05, 0.15)$; 3) varying correlations and magnitudes of functional measurement error, $l_{U} \in (0.05,0.15,0.25)$, and $\sigma_{U_1} \in (1.5, 2, 3)$ with fixed measurement error in the scalar covariate; and 4) different structures for the variance-covariance matrix of the true functional covariate and measurement error including compound symmetric, autoregressive one, squared exponential, and unstructured. For the last set of simulations, we set $n=2,000$, $\sigma_X= 4$, $\sigma_U= 2$, $l_X = 0.25$, $l_U=0.25$, $\rho_{X_1(t)}=0.25$, $\rho_{U_1(t)}=0.25$,  and $\sigma_{U} = 2$. The $l_X$ and $l_U$ are the tuning parameters in the exponential covariance function of the functional true covariate, $X_1(t)$, and measurement error $U_1(t)$, respectively. The $\rho_{X_1(t)}$ and $\rho_{U_1(t)}$ are the correlations between adjacent locations in the compound symmetric and autoregressive one correlation matrix of the functional true covariate, $X_1(t)$ and measurement error $U_1(t)$, respectively. We generated the unstructured covariance matrix with a correlation between adjacent locations in the covariance matrix, $\rho$, to control the overall correlation strength. We generated the correlation between two adjacent locations randomly from a uniform distribution with minimum value as $\max\{0, \rho_{X_1(t)}-0.25\}$ and maximum value as $\min\{\rho_{X_1(t)}+0.25, 1\}$ for $X_1(t)$. By tuning the value of $\rho_{X_1(t)}$, we control the overall correlation strength of the unstructured covariance matrix associated with $X_1(t)$. Similarly, the correlation between two adjacent locations in the unstructured correlation matrix of $U_1(t)$ is between $\max\{0, \rho_{U_1(t)}-0.25\}$ and $\min\{\rho_{U_1(t)}+0.25, 1\}$, allowing control of the overall correlation strength of the unstructured covariance matrix associated with $U_1(t)$ in the simulated data.

Each of the four sets of simulations included $500$ replications. We obtained the number of basis functions, $K_n$, as a function of the sample size n as we previously described in \cite{tekwe2022estimation}.

We considered five estimation methods: oracle, 
average, naive, SIMEX, and RC. We obtained the oracle estimator from the true measures of both the functional covariate {$X_{1i}(t)$} and the scalar covariate ($X_{2i}$). The average estimator was the average of observed measurements across all the repeated sessions {$\overline{W_{1i}}(t)$ and $\overline{W_{2i}}$}. We obtained the naive estimator with the observed measures from a single session $W_{1ij}(t)$ and $W_{2ij}$, selecting the observations from the first repeated session, $W_{1i1}(t)$ and $W_{2i1}$. We obtained the SIMEX and RC estimators with the SIMEX and RC methods, respectively.

 Let $\bar{\beta}_1(t) = \frac{1}{500}\sum_r^{500}\widehat{\beta}_{1r}(t)$ and $\bar{\beta_2} = \frac{1}{500}\sum_r^{500} \widehat{\beta}_{2r}$, where $\widehat{\beta_{1r}}(t)$ and  $\widehat{\beta}_{2r}$  are the estimators from the $r$-th replication. We use the following measures for comparisons of the estimators for ${\beta}_1(t)$ and ${\beta_2}$. In particular, we assessed the performance of $\widehat{\beta}_1(t)$ with the average squared bias (ABias$^2$), the average variance (Avar), and the average integrated mean squared error (AIMSE):
\begin{eqnarray*}
ABias^2\{\widehat{\beta_1}(t)\} &=& \frac{1}{t_{grid}} \sum_{h=1}^{t_{grid}} \left\{\bar{\beta_1}(t_{h}) - \beta_1(t_{h})\right\}^2, \\
AVar\{\widehat{\beta}_{1}(t)\} & =& \frac{1}{t_{grid}} \sum_{h=1}^{t_{grid}} \frac{1}{500} \sum_{r=1}^{500}\left\{ \widehat{\beta}_{1r}(t_{h})-\bar{\beta}_1(t_{h}) \right\}^2, \text{and}\\
AIMSE\{\widehat{\beta}_1(t)\} & =& ABias^2\{\widehat{\beta}_1(t)\} + Avar\{\widehat{\beta}_1(t)\},
\end{eqnarray*}
in which $\{t_h\}^{t_{grid}}_{h=1}$ is a sequence of equally spaced grid points in $[0,1]$.
We also assessed the performance of $\widehat{\beta}_2$ with the  squared bias (Bias$^2$), the variance (Var), and the  mean squared error (MSE):
\begin{eqnarray*}
Bias^2(\widehat{\beta_2}) &=& (\bar{\beta_2} - \beta_2)^2, \\
Var(\widehat{\beta_2}) &=& \frac{1}{500} \sum_{r=1}^{500} (\widehat{\beta_{2r}} -\bar{\beta_2})^2, \text{and}\\
MSE(\widehat{\beta_2}) &=& Bias^2(\widehat{\beta_2}) + Var(\widehat{\beta_2}).
\end{eqnarray*}

When $\sigma_{U_1}$ and $\sigma_{U_2}$ are the same, we use $\sigma_{U}$ to denote their common value. Similarly, we use $\sigma_{X}$ to denote the same standard deviation for covariates when $\sigma_{X_1}$ and $\sigma_{X_2}$ are equal. Otherwise, we denote the values as $\sigma_{U_1}$, $\sigma_{U_2}$, $\sigma_{X_1}$, and $\sigma_{X_2}$.

\subsection{Simulation results}
\subsubsection{Effects of varying sample sizes}
Tables \ref{Table:samplesize_function} and \ref{Table:samplesize_scalar} show the results of the impacts of varying sample sizes on the estimators for functional and scalar covariates, respectively. For the functional covariate, as the sample size increased, the ABias$^2$ of the oracle, SIMEX, and RC estimators consistently decreased, but the average and naive-based estimators changed little, with notable biases even when the sample size was large ($n=5,000$). For the scalar covariate, Bias$^2$ increased as sample size increased for the oracle, RC, and average estimators, and also for the SIMEX estimator when measurement error was high. When measurement error was low, the Bias$^2$ in the SIMEX estimator decreased with increasing sample size. Despite the contrasting patterns for the SIMEX estimator under different levels of measurement error, Bias$^2$ was notably smaller for the SIMEX estimator than for the RC estimator in all conditions except with a sample size of $500$ and moderate error ($\sigma_U$= $1.5$). Bias$^2$ was consistently high for the naive estimator in all conditions. For both functional and scalar covariates, in terms of bias, the oracle estimator performed best, followed in order by the SIMEX, RC, average, and naive estimators. 

The AVar/Var of all the estimators decreased with increasing sample size for both the scalar and functional covariates. Across conditions, the naive estimator had the lowest AVar/Var, followed in ascending order by the average, RC, and SIMEX estimators. The AVar/Var of the oracle estimator were similar to those of the RC estimator. The AIMSE/MSE decreased with increasing sample size for the oracle and SIMEX estimators in all measurement error conditions as well as for the RC and average estimators for the functional covariate and under relatively low measurement error for the scalar covariate. The AIMSE/MSE did not vary meaningfully across sample sizes for the naive estimator and increased somewhat with sample size for the average estimator under relatively high measurement error in the scalar covariate.  For the functional covariate, the oracle, SIMEX, RC, and average estimators had generally similar AIMSE values that were substantially lower than that of the naive estimator for a given sample size-measurement error combination. The simulations showed the same pattern for MSE in the scalar covariate except that  the average estimator performed worse than the oracle, SIMEX, and RC estimators under relatively high measurement error. The SIMEX, RC, average, and naive estimators are all approximately consistent \cite{carroll2006measurement}. Thus, there is a trade-off between variance and bias in estimation, consistent with the naive estimator's lowest average variance/variance and highest average bias/bias in Tables \ref{Table:samplesize_function} and \ref{Table:samplesize_scalar}.

\begin{table}[h] 
\caption{The effect of varying sample sizes on the performance of $\widehat{\beta_1}(t)$ with the oracle, SIMEX, RC, average, and naive estimators, as indicated by the average squared bias (ABias$^2$), average sample variance (AVar), and average integrated mean squared error (AIMSE) for $n \in (500,1000, 2000, 5000)$; $\sigma_{X_1}= \sigma_{X_2}=\sigma_{X}=3$; and $\sigma_{U_1}=\sigma_{U_2}= \sigma_{U}= 1.5$ or $3$. 
\label{Table:samplesize_function}}
\centering
\scalebox{0.78}{
\begin{tabular}{r rrrrr|rrrr r|rr rrr}
		\hline
			\multicolumn{16}{c} {\textbf{$\sigma_{U} = 1.5$}}\\
		\hline
		\multicolumn{1}{c} {} &\multicolumn{5}{c} {ABias$^2$} &\multicolumn{5}{c} {AVar} &\multicolumn{5}{c} {AIMSE} \\
		\hline
		n &  Oracle & SIMEX & RC & Average & Naive  & Oracle & SIMEX & RC& Average & Naive  & Oracle & SIMEX& RC & Average & Naive  \\

		\hline
 500 & 0.0058 & 0.0059 & 0.0110 & 0.0157 & 0.1097 & 0.1779 & 0.1963 & 0.1707 & 0.1546 & 0.1063 & 0.1836 & 0.2023 & 0.1818 & 0.1702 & 0.2160\\

1000 & 0.0058 & 0.0061 & 0.0122 & 0.0172 & 0.1138 & 0.0842 & 0.0924 & 0.0805 & 0.0730 & 0.0502 & 0.0900 & 0.0985 & 0.0927 & 0.0902 & 0.1640\\

2000 & 0.0002 & 0.0004 & 0.0069 & 0.0124 & 0.1122 & 0.0541 & 0.0598 & 0.0541 & 0.0491 & 0.0369 & 0.0543 & 0.0602 & 0.0610 & 0.0615 & 0.1491\\

5000 & 0.0002 & 0.0003 & 0.0062 & 0.0114 & 0.1099 & 0.0261 & 0.0289 & 0.0268 & 0.0243 & 0.0185 & 0.0262 & 0.0292 & 0.0330 & 0.0358 & 0.1284\\
\hline
			\multicolumn{16}{c} {\textbf{$\sigma_{U} = 3$}}\\
		\hline
		\multicolumn{1}{c} {} &\multicolumn{5}{c} {ABias$^2$} &\multicolumn{5}{c} {AVar} &\multicolumn{5}{c} {AIMSE} \\
		\hline
		n & Oracle & SIMEX & RC & Average & Naive  & Oracle & SIMEX & RC& Average & Naive  & Oracle & SIMEX& RC & Average & Naive  \\

		\hline
500 & 0.0058 & 0.0076 & 0.0151 & 0.0432 & 0.2198 & 0.1779 & 0.2036 & 0.1868 & 0.1291 & 0.0712 & 0.1836 & 0.2111 & 0.2019 & 0.1723 & 0.2910\\

1000 & 0.0058 & 0.0074 & 0.0148 & 0.0424 & 0.2191 & 0.0842 & 0.1010 & 0.0922 & 0.0638 & 0.0339 & 0.0900 & 0.1084 & 0.1071 & 0.1062 & 0.2529\\

2000 & 0.0002 & 0.0039 & 0.0130 & 0.0438 & 0.2248 & 0.0541 & 0.0629 & 0.0610 & 0.0423 & 0.0251 & 0.0543 & 0.0668 & 0.0740 & 0.0861 & 0.2499\\

5000 & 0.0002 & 0.0033 & 0.0120 & 0.0421 & 0.2229 & 0.0261 & 0.0308 & 0.0307 & 0.0213 & 0.0122 & 0.0262 & 0.0341 & 0.0428 & 0.0634 & 0.2350\\
\hline
\end{tabular}
}
\end{table}

\begin{table}[h] 
\caption{The effect of varying sample sizes on the performance of $\widehat{\beta_2}$ with the oracle, SIMEX, RC, average, and naive estimators, as indicated by the average squared bias (Bias$^2$), average sample variance (Var), and mean squared error (MSE) for  $n \in (500,1000,2000,5000)$;  $\sigma_{X_1}=\sigma_{X_2}=\sigma_{X}= 3$; and $\sigma_{U_1}=\sigma_{U_2}=\sigma_{U} = 1.5$ or $3$. 
\label{Table:samplesize_scalar}}
\centering
\scalebox{0.78}{
\begin{tabular}{r rrrrr|rrrr r|rr rrr}
		\hline
	\multicolumn{16}{c} {\textbf{$\sigma_{U} = 1.5$}}\\
		\hline
		\multicolumn{1}{c} {} &\multicolumn{5}{c} {Bias$^2$} &\multicolumn{5}{c} {Var} &\multicolumn{5}{c} {MSE} \\
		\hline
		n & Oracle & SIMEX & RC & Average & Naive  & Oracle & SIMEX & RC& Average & Naive  & Oracle & SIMEX& RC & Average & Naive  \\
		\hline
500 & 0.0055 & 0.0046 & 0.0000 & 0.0021 & 0.1136 & 0.0261 & 0.0320 & 0.0219 & 0.0199 & 0.0085 & 0.0315 & 0.0366 & 0.0219 & 0.0220 & 0.1220\\

 1000 & 0.0011 & 0.0008 & 0.0011 & 0.0062 & 0.1216 & 0.0108 & 0.0138 & 0.0095 & 0.0086 & 0.0042 & 0.0119 & 0.0146 & 0.0105 & 0.0148 & 0.1258\\

2000 & 0.0004 & 0.0001 & 0.0024 & 0.0089 & 0.1309 & 0.0043 & 0.0055 & 0.0039 & 0.0035 & 0.0018 & 0.0047 & 0.0056 & 0.0063 & 0.0125 & 0.1327\\

 5000 & 0.0001 & 0.0000 & 0.0032 & 0.0104 & 0.1321 & 0.0019 & 0.0022 & 0.0015 & 0.0014 & 0.0007 & 0.0019 & 0.0022 & 0.0047 & 0.0118 & 0.1327\\
\hline
	\multicolumn{16}{c} {\textbf{$\sigma_{U} = 3$}}\\
		\hline
		\multicolumn{1}{c} {} &\multicolumn{5}{c} {Bias$^2$} &\multicolumn{5}{c} {Var} &\multicolumn{5}{c} {MSE} \\
		\hline
		n & Oracle & SIMEX & RC & Average & Naive  & Oracle & SIMEX & RC& Average & Naive  & Oracle & SIMEX& RC & Average & Naive  \\
		\hline
 500 & 0.0055 & 0.0020 & 0.0162 & 0.0753 & 0.4503 & 0.0261 & 0.0267 & 0.0152 & 0.0105 & 0.0025 & 0.0315 & 0.0287 & 0.0315 & 0.0858 & 0.4529\\

 1000 & 0.0011 & 0.0055 & 0.0237 & 0.0873 & 0.4544 & 0.0108 & 0.0121 & 0.0070 & 0.0049 & 0.0013 & 0.0119 & 0.0176 & 0.0307 & 0.0922 & 0.4557\\

 2000 & 0.0004 & 0.0088 & 0.0286 & 0.0949 & 0.4640 & 0.0043 & 0.0049 & 0.0029 & 0.0020 & 0.0006 & 0.0047 & 0.0137 & 0.0315 & 0.0969 & 0.4646\\

 5000 & 0.0001 & 0.0098 & 0.0301 & 0.0973 & 0.4631 & 0.0019 & 0.0019 & 0.0011 & 0.0008 & 0.0002 & 0.0019 & 0.0117 & 0.0312 & 0.0981 & 0.4633\\
\hline
\end{tabular}
}
\end{table}

\subsubsection{Impacts of varying magnitudes of measurement error}
\begin{figure}[h]
	\centering
	\includegraphics[width=18cm]{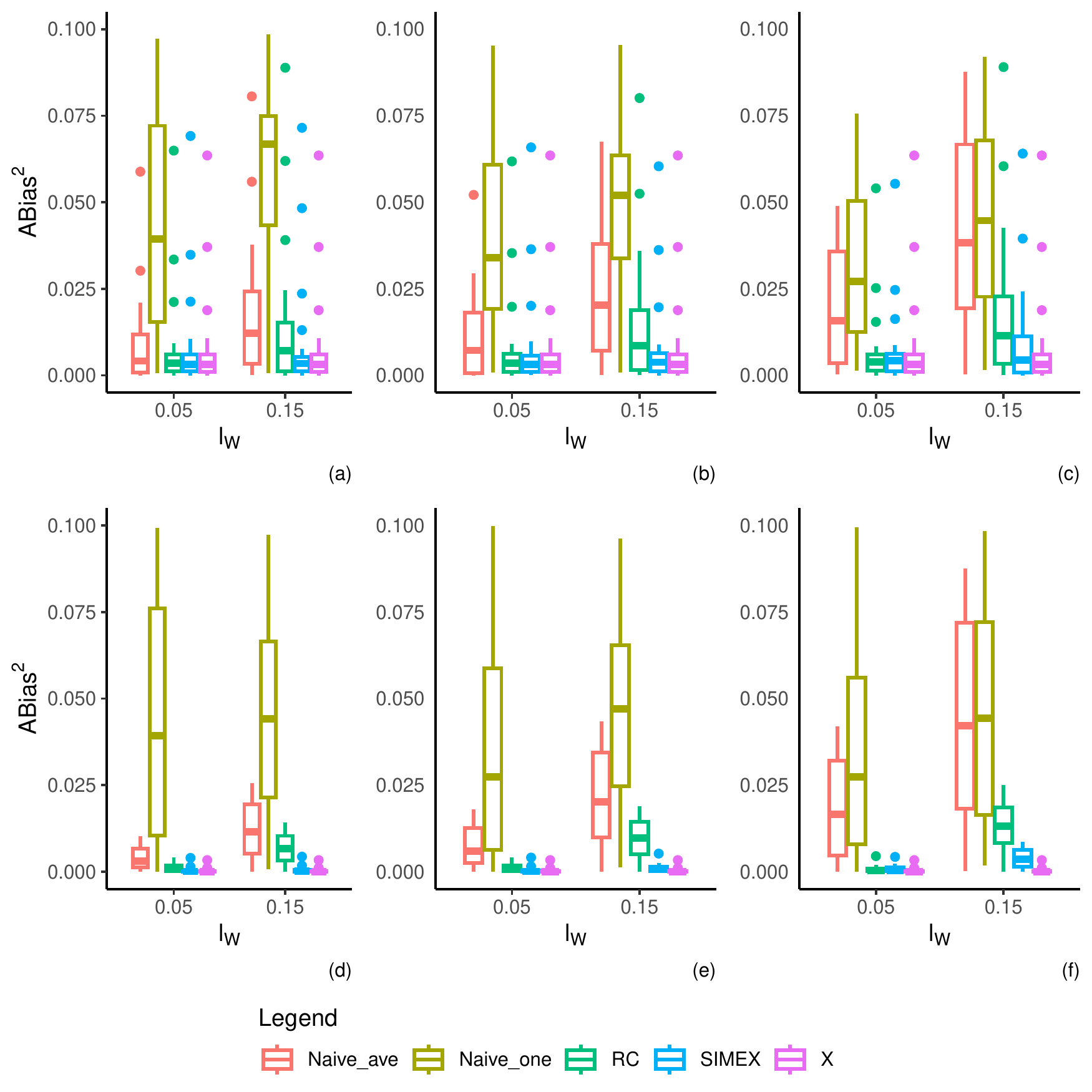}
	\caption{Boxplots of average squared biases (ABias$^2$) of $\widehat{\beta_1}(t)$ obtained under the Average, Naive, SIMEX, RC, and Oracle methods. Each plot compares the boxplot of ABias$^2$ from the  Average, Naive,  SIMEX, RC, and Oracle methods estimators of $\widehat{\beta_1}(t)$ in the generalized functional linear function when  $\sigma_{X_1}=\sigma_{X_2}=3$ and $l_X=0.05$. The top three figures correspond to $n=500$, with (a) $\sigma_{U_1}=1.5$, (b) $\sigma_{U_1}=2$, and (c) $\sigma_{U_1}=3$ for $l_U=0.05,0.15$. The bottom three figures correspond to $n=2000$ with, (d) $\sigma_{U_1}=1.5$, (e) $\sigma_{U_1}=2$, and (f) $\sigma_{U_1}=3$ for $l_U=0.05,0.15$. $l_X$ and $l_U$ are the tuning parameters in the exponential covariance function of the functional true covariate and measurement error, respectively. \label{boxplot_bias}}
\end{figure}

\begin{figure}[h]
	\centering
	\includegraphics[width=18cm]{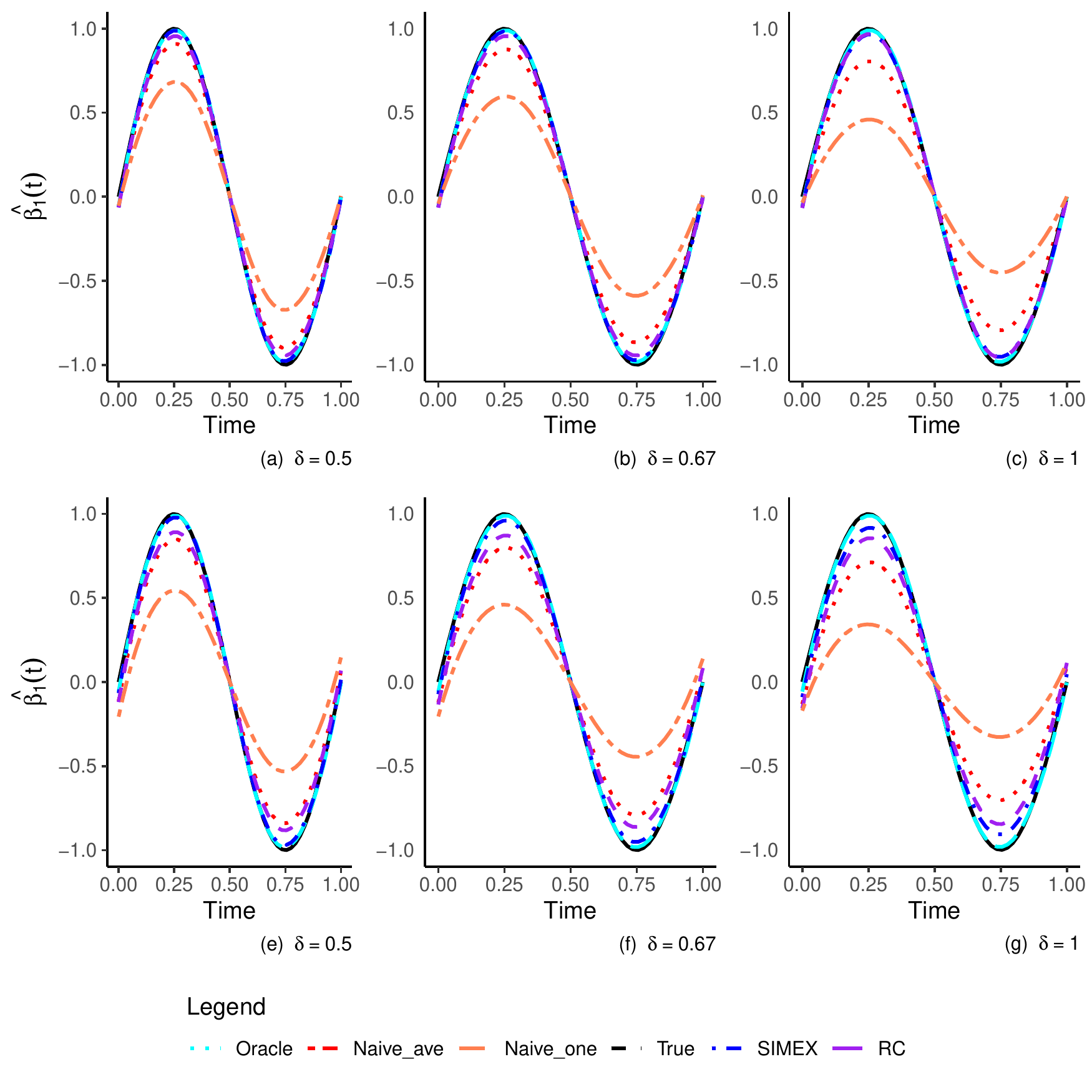}
	\caption{Plots of the estimated regression coefficient function in the generalized functional linear function from simulations study when $n=2000$, $\sigma_{X_1}=\sigma_{X_2}=3$, and $l_X=0.05$, with (a) $\sigma_{U_1}=1.5,l_U=0.05$; (b) $\sigma_{U_1}=2,l_U=0.05$; (c), $\sigma_{U_1}=3,l_U=0.05$; (d), $\sigma_{U_1}=1.5,l_U=0.15$; (e), $\sigma_{U_1}=2,l_U=0.15$; and (f), $\sigma_{U_1}=3,l_U=0.15$. $\delta=\frac{\sigma_{U_1}}{\sigma_{X_1}}$. In each plot, the solid black line represents the true coefficient function, while the cyan-long-dashed, coral-dotted, red two-dashed, purple-dashed, and blue-dot-dashed lines represent the averaged coefficient estimators over 500 replicates from the Oracle, Average, Naive, RC, and SIMEX methods, respectively. The terms, $l_X$ and $l_U$, are the tuning parameters in the exponential covariance function of the functional true covariate and measurement error, respectively.\label{sim_beta_plot}}
\end{figure}

Tables \ref{Table:MEsize_func} and \ref{Table:MEsize_scalar} and Figures \ref{boxplot_bias} - \ref{barplot} show the results on the impacts of varying degrees of measurement error on the estimators for functional and scalar covariates, respectively. The oracle estimator consistently had the smallest bias that was constant across levels of measurement error. For all of the other estimators, bias usually increased with increasing levels of measurement error. Apart from the oracle estimator, the SIMEX estimator had the lowest bias, followed in ascending order by the RC, average, and naive estimators. As measurement error increased, the comparative advantage of the SIMEX estimator over the RC estimator tended to decline somewhat. 

The oracle estimator had the smallest AVar/Var that were constant across levels of measurement error. As measurement error increased, the AVar of the SIMEX and RC estimators increased somewhat for the functional covariate, but the Var for these estimators decreased somewhat for the scalar covariate. For both the functional and scalar covariates, the AVar/Var declined for the average and naive estimators as measurement error increased. Aside from the oracle estimator, the AIMSE/MSE rose with increasing measurement error for all estimators, with the SIMEX estimator generally performing the best, followed in order by the RC, average, and naive estimators.

Figure \ref{barplot} shows the bias of $\widehat{\beta_2}$ for the different estimators under different magnitudes of measurement error. As indicated by the results for the oracle estimator, the bias of the $\widehat{\beta_2}$ was positive when there was no measurement error. The biases for the naive and average estimators were negative and became increasingly negative as $\sigma_{W_2}$ increased. Therefore, the bias of $\widehat{\beta_2}$ is negative when there are uncorrected measurement errors, leading to attenuated estimates of the association between the scalar covariate and the outcome. As $\sigma_{W_2}$ increased, the bias of the measurement error-corrected estimators (SIMEX and RC) changed from slightly positive to slightly to moderately negative. This indicates that the ability of SIMEX and RC to correct measurement errors declines as measurement error increases, with the RC estimator more sensitive in this way than the SIMEX estimator.

\begin{table}[h] 
\caption{The effect of varying magnitudes of measurement error on the performance of $\widehat{\beta_1}(t)$ with the oracle, SIMEX, RC, average, and naive estimators, as indicated by the average squared bias (ABias$^2$), average sample variance (Avar), and average integrated mean squared error (AIMSE) for $n=2,000$;  $\sigma_{X_1}=\sigma_{X_2}=\sigma_{X}=3$; $\sigma_{U_1} =\sigma_{U_2}=\sigma_{U} \in(1.5,2,3,4)$; $l_X =0.05$; and $l_U =0.15$, where $l_X$ and $l_U$ are the tuning parameters in the exponential covariance function of the functional true covariate and measurement error, respectively.
\label{Table:MEsize_func}}
\centering
\scalebox{0.78}{
\begin{tabular}{r rrrr r|rr rrr|r rrrr}
		\hline
		\multicolumn{1}{c} {} &\multicolumn{5}{c} {ABias$^2$} &\multicolumn{5}{c} {Avar} &\multicolumn{5}{c} {AIMSE} \\
		\hline
		$\sigma_{U}$ & Oracle & SIMEX & RC & Average & Naive  & Oracle & SIMEX & RC& Average & Naive  & Oracle & SIMEX& RC & Average & Naive  \\
		\hline
1.5 & 0.0002 & 0.0004 & 0.0069 & 0.0124 & 0.1122 & 0.0541 & 0.0598 & 0.0541 & 0.0491 & 0.0369 & 0.0543 & 0.0602 & 0.0610 & 0.0615 & 0.1491\\

2.0 & 0.0002 & 0.0009 & 0.0097 & 0.0215 & 0.1555 & 0.0541 & 0.0608 & 0.0553 & 0.0466 & 0.0326 & 0.0543 & 0.0617 & 0.0651 & 0.0680 & 0.1881\\

3.0 & 0.0002 & 0.0039 & 0.0130 & 0.0438 & 0.2248 & 0.0541 & 0.0629 & 0.0610 & 0.0423 & 0.0251 & 0.0543 & 0.0668 & 0.0740 & 0.0861 & 0.2499\\

4.0 & 0.0002 & 0.0103 & 0.0129 & 0.0697 & 0.2751 & 0.0541 & 0.0643 & 0.0705 & 0.0382 & 0.0194 & 0.0543 & 0.0746 & 0.0834 & 0.1078 & 0.2945\\
\hline	
\end{tabular}
}
\end{table}

\begin{table}[h] 

\caption{The effect of varying magnitudes of measurement error on the performance of $\widehat{\beta_2}$ with the oracle, SIMEX, RC, average, and naive estimators, as indicated by the squared bias (Bias$^2$), sample variance (Var), and mean squared error (MSE) for $n=2,000$; $\sigma_{X_1}=\sigma_{X_2}=\sigma_{X}=3$; and $\sigma_{U_1} =\sigma_{U_2}=\sigma_{U} \in (1.5,2,3,$ or $4)$.
\label{Table:MEsize_scalar}}
\centering
\scalebox{0.78}{
\begin{tabular}{r rrrrr|rrrr r|rr rrr}
		\hline
		\multicolumn{1}{c} {} &\multicolumn{5}{c} {Bias$^2$} &\multicolumn{5}{c} {Var} &\multicolumn{5}{c} {MSE} \\
		\hline
		$\sigma_{U}$ & Oracle & SIMEX & RC & Average & Naive  & Oracle & SIMEX & RC& Average & Naive  & Oracle & SIMEX& RC & Average & Naive  \\

\hline


1.5 & 0.0011 & 0.0008 & 0.0011 & 0.0062 & 0.1216 & 0.0108 & 0.0138 & 0.0095 & 0.0086 & 0.0042 & 0.0119 & 0.0146 & 0.0105 & 0.0148 & 0.1258\\

2.0 & 0.0011 & 0.0001 & 0.0051 & 0.0219 & 0.2359 & 0.0108 & 0.0139 & 0.0085 & 0.0072 & 0.0028 & 0.0119 & 0.0140 & 0.0137 & 0.0291 & 0.2387\\

3.0 & 0.0011 & 0.0055 & 0.0237 & 0.0873 & 0.4544 & 0.0108 & 0.0121 & 0.0070 & 0.0049 & 0.0013 & 0.0119 & 0.0177 & 0.0307 & 0.0922 & 0.4557\\

4.0 & 0.0011 & 0.0369 & 0.0504 & 0.1838 & 0.6129 & 0.0108 & 0.0091 & 0.0061 & 0.0032 & 0.0007 & 0.0119 & 0.0460 & 0.0564 & 0.1870 & 0.6136\\
\hline




\end{tabular}
}
\end{table}

\subsubsection{Impacts of the functional measurement error on the scalar covariate}

Varying correlations and magnitudes of the functional measurement error had essentially no impact on the Bias$^2$, Var, and MSE of $\widehat{\beta_2}$ for each of the five estimators (Table \ref{Table:func_on_scalar}). Although Table \ref{Table:func_on_scalar} shows the simulation results for $n=1,000$ only, we observed similar results for $n \in (500, 2000, 5000)$.

\begin{table}[h] 

\caption{The effect of varying correlations and magnitudes of functional measurement error on the performance of $\widehat{\beta_2}$ with the oracle, SIMEX, RC, average, and naive estimators, as indicated by the squared bias (Bias$^2$), sample variance (Var), and mean squared error (MSE) for $n=1,000$; $\sigma_{X_1}= \sigma_{X_2}=3$; $\sigma_{U_2} =2$; $\sigma_{U_1} =1.5,2,$ or $3$; $l_X=0.05$; and $l_U =0.05,0.15,$ or $0.25$, where $l_X$ and $l_U$ are the tuning parameters in the exponential covariance function of the functional true covariate and measurement error, respectively.\label{Table:func_on_scalar}}
\centering
\scalebox{0.78}{
\begin{tabular}{r rrrrr|rrrr r|rr rrr}
		\hline
			\multicolumn{16}{c} {\textbf{$l_{U}=0.05$}}\\
		\hline
		\multicolumn{1}{c} {} &\multicolumn{5}{c} {Bias$^2$} &\multicolumn{5}{c} {Var} &\multicolumn{5}{c} {MSE} \\
		\hline
		$\sigma_{U_1}$& Oracle & SIMEX & RC & Average & Naive  & Oracle & SIMEX & RC& Average & Naive  & Oracle & SIMEX& RC & Average & Naive  \\

\hline
1.5 & 0.0009 & 0.0001 & 0.0050 & 0.0216 & 0.2323 & 0.0104 & 0.0130 & 0.0081 & 0.0069 & 0.0027 & 0.0113 & 0.0131 & 0.0131 & 0.0284 & 0.2350\\

2.0 & 0.0009 & 0.0001 & 0.0051 & 0.0218 & 0.2333 & 0.0104 & 0.0130 & 0.0080 & 0.0068 & 0.0027 & 0.0113 & 0.0130 & 0.0131 & 0.0286 & 0.2360\\

3.0 & 0.0009 & 0.0000 & 0.0054 & 0.0223 & 0.2345 & 0.0104 & 0.0129 & 0.0079 & 0.0067 & 0.0027 & 0.0113 & 0.0129 & 0.0133 & 0.0291 & 0.2372\\
\hline
	\multicolumn{16}{c} {\textbf{$l_{U}=0.15$}}\\
		\hline
		\multicolumn{1}{c} {} &\multicolumn{5}{c} {Bias$^2$} &\multicolumn{5}{c} {Var} &\multicolumn{5}{c} {MSE} \\
		\hline
		$\sigma_{U_1}$ & Oracle & SIMEX & RC & Average & Naive  & Oracle & SIMEX & RC& Average & Naive  & Oracle & SIMEX& RC & Average & Naive  \\
\hline
1.5 & 0.0009 & 0.0000 & 0.0053 & 0.0222 & 0.2344 & 0.0104 & 0.0134 & 0.0082 & 0.0069 & 0.0026 & 0.0113 & 0.0134 & 0.0135 & 0.0291 & 0.2371\\

2.0 & 0.0009 & 0.0000 & 0.0055 & 0.0226 & 0.2354 & 0.0104 & 0.0128 & 0.0079 & 0.0067 & 0.0026 & 0.0113 & 0.0128 & 0.0134 & 0.0293 & 0.2380\\

3.0 & 0.0009 & 0.0000 & 0.0059 & 0.0233 & 0.2366 & 0.0104 & 0.0125 & 0.0078 & 0.0066 & 0.0026 & 0.0113 & 0.0126 & 0.0137 & 0.0299 & 0.2392\\
\hline
	\multicolumn{16}{c} {\textbf{$l_{U}=0.25$}}\\
		\hline
		\multicolumn{1}{c} {} &\multicolumn{5}{c} {Bias$^2$} &\multicolumn{5}{c} {Var} &\multicolumn{5}{c} {MSE} \\
		\hline
		$\sigma_{U_1}$ & Oracle & SIMEX & RC & Average & Naive  & Oracle & SIMEX & RC& Average & Naive  & Oracle & SIMEX& RC & Average & Naive  \\

\hline
1.5 & 0.0009 & 0.0000 & 0.0054 & 0.0224 & 0.2342 & 0.0104 & 0.0130 & 0.0080 & 0.0068 & 0.0027 & 0.0113 & 0.0130 & 0.0134 & 0.0292 & 0.2369\\

2.0 & 0.0009 & 0.0000 & 0.0056 & 0.0227 & 0.2350 & 0.0104 & 0.0128 & 0.0080 & 0.0068 & 0.0027 & 0.0113 & 0.0128 & 0.0136 & 0.0295 & 0.2377\\

3.0 & 0.0009 & 0.0000 & 0.0060 & 0.0234 & 0.2359 & 0.0104 & 0.0127 & 0.0079 & 0.0067 & 0.0027 & 0.0113 & 0.0127 & 0.0138 & 0.0300 & 0.2386\\
\hline
\end{tabular}
}
\end{table}

\subsubsection{Impacts of varying functional measurement error variance-covariance matrix structures} \label{sim_l}

The naive, average, and RC estimators are based on the assumption that the measurement error and true measurements have the same variance-covariance structure. The naive and average estimators treat the observed measurements as true measurements. The RC estimator involves FUI to approximate the true measurement with the observed measurement at each grid location (time point) discretely and does not address the correlation between observed measurements at adjacent locations along the functional continuum. Thus, the RC approximation of the true functional measurements  naturally has a very similar correlation structure as the observed measurement. The SIMEX estimator does not involve any assumptions on the variance-covariance structure of the measurement error or true measurement. Instead, we estimate measurement error variance-covariance matrix with repeated measures, which distinguish the variance-covariance matrix of measurement error from that of the true measurement. Also, by extrapolating back to the "measurement error-free" situation in the extrapolation step of SIMEX, the expected measurement error variance-covariance is 0 and thus has a very limited effect on the estimated coefficients. 

Table \ref{Table:correlation} shows that, across combinations of correlation structures of $X_1(t)$ and $U_1(t)$, the ABias$^2$ of the SIMEX estimator was small and similar to that of the oracle estimator. The RC estimator tended to have the next smallest ABias$^2$, followed in ascending order by the average and naive estimators. However, the AVar was smallest for the naive estimator, followed in order by the average, RC, oracle, and SIMEX estimators. The ranges of AIMSE values across estimators were narrow, with the average estimator having the lowest values and the SIMEX generally having the highest. 
 
\begin{table}[h]
\caption{The effect of varying functional measurement error variance-covariance matrix structures on the performance of $\widehat{\beta_1}(t)$ with the oracle, SIMEX, RC, average, and naive estimators. Results are for combinations of unstructured (UN), compound symmetric (CS), autoregressive one (AR1), and squared exponential (EXP) structures. Performance indicated by the average squared bias (ABias$^2$), average sample variance (Avar), and average integrated mean squared error (AIMSE) for $n=2,000$, $\sigma_X= 4$, $\sigma_U= 2$, $l_X = 0.25$, $l_U=0.25$, $\rho_{X_1(t)}=0.25$, $\rho_{U_1(t)}=0.25$, and $\sigma_{U} = 2$. The $l_X$ and $l_U$ are the tuning parameters in the exponential covariance function of the functional true covariate and measurement error, respectively. The $\rho_{X_1(t)}$ and $\rho_{U_1(t)}$ are the correlations between adjacent locations in the compound symmetric and autoregressive one correlation matrix of the functional true covariate and measurement error, respectively. The scale of $\rho_{X_1(t)}$ and $\rho_{U_1(t)}$ roughly represent the strengths of correlations between different locations in the unstructured correlation matrix of the functional true covariate and measurement error, respectively.
\label{Table:correlation}}
\centering
\scalebox{0.73}{
	\begin{tabular}{ll| lllll|lllll|lllll}
	\hline
		\multicolumn{17}{c} {$\boldsymbol{\rho_{X_1(t)}}=0.25, \quad \boldsymbol{\rho_{U_1(t)}}=0.25$}\\
		\hline
		\multicolumn{2}{c} {} &\multicolumn{5}{c} {ABias$^2$} &\multicolumn{5}{c} {AVar} &\multicolumn{5}{c} {AIMSE} \\
		\hline
		Cov$_{X_1(t)}$ &Cov$_{U_1(t)}$ & Oracle & SIMEX & RC & Average & Naive  & Oracle & SIMEX & RC& Average & Naive  & Oracle & SIMEX& RC & Average & Naive  \\
		\hline
AR1 & AR1 & 0.0005 & 0.0005 & 0.0049 & 0.0099 & 0.0964 & 0.1303 & 0.1518 & 0.1237 & 0.1122 & 0.076 & 0.1308 & 0.1523 & 0.1286 & 0.122 & 0.1724\\

AR1 & CS & 0.0005 & 0.0004 & 0.0026 & 0.0066 & 0.0666 & 0.1303 & 0.1511 & 0.1217 & 0.1104 & 0.0739 & 0.1308 & 0.1515 & 0.1244 & 0.1169 & 0.1405\\

AR1 & UN & 0.0002 & 0.0005 & 0.0035 & 0.0087 & 0.0734 & 0.1293 & 0.1513 & 0.1226 & 0.1097 & 0.0733 & 0.1295 & 0.1518 & 0.1261 & 0.1184 & 0.1467\\

CS & AR1 & 0.0003 & 0.0007 & 0.0102 & 0.0168 & 0.1445 & 0.2253 & 0.2573 & 0.1952 & 0.1768 & 0.1038 & 0.2256 & 0.258 & 0.2055 & 0.1936 & 0.2483\\

CS & CS & 0.0003 & 0.0006 & 0.0049 & 0.0098 & 0.0947 & 0.2253 & 0.2582 & 0.2129 & 0.1929 & 0.1291 & 0.2256 & 0.2588 & 0.2178 & 0.2028 & 0.2239\\

CS & UN & 0.0002 & 0.0004 & 0.0054 & 0.0115 & 0.0982 & 0.2207 & 0.2556 & 0.2107 & 0.1884 & 0.1312 & 0.2209 & 0.256 & 0.2161 & 0.1999 & 0.2294\\

UN & AR1 & 0.0005 & 0.0016 & 0.0143 & 0.0207 & 0.1469 & 0.2205 & 0.2609 & 0.1934 & 0.1772 & 0.1057 & 0.221 & 0.2625 & 0.2077 & 0.198 & 0.2526\\

UN & CS & 0.0005 & 0.001 & 0.0075 & 0.0126 & 0.095 & 0.2205 & 0.2531 & 0.2047 & 0.1876 & 0.1298 & 0.221 & 0.2542 & 0.2122 & 0.2002 & 0.2248\\

UN & UN & 0.0008 & 0.0005 & 0.0044 & 0.0093 & 0.0978 & 0.2477 & 0.2817 & 0.2272 & 0.2059 & 0.1268 & 0.2485 & 0.2822 & 0.2316 & 0.2151 & 0.2246\\
\hline
		\multicolumn{17}{c} {$\boldsymbol{\rho_{X_1(t)}}=0.25, \quad \boldsymbol{l_{U}}=0.25$}\\
		\hline
		\multicolumn{2}{c} {} &\multicolumn{5}{c} {ABias$^2$} &\multicolumn{5}{c} {AVar} &\multicolumn{5}{c} {AIMSE} \\
		\hline
		Cov$_{X_1(t)}$ &Cov$_{U_1(t)}$ & Oracle & SIMEX & RC & Average & Naive  & Oracle & SIMEX & RC& Average & Naive  & Oracle & SIMEX& RC & Average & Naive  \\
		\hline
AR1 & EXP & 0.0005 & 0.0009 & 0.0144 & 0.0218 & 0.1652 & 0.113 & 0.1355 & 0.1014 & 0.0919 & 0.0514 & 0.1135 & 0.1365 & 0.1159 & 0.1137 & 0.2166\\

CS & EXP & 0.0005 & 0.0006 & 0.0023 & 0.0059 & 0.0504 & 0.113 & 0.1255 & 0.1035 & 0.0938 & 0.0682 & 0.1135 & 0.1261 & 0.1058 & 0.0998 & 0.1186\\

UN & EXP & 0.0004 & 0.0006 & 0.002 & 0.007 & 0.0588 & 0.1074 & 0.1203 & 0.0997 & 0.0878 & 0.0613 & 0.1078 & 0.121 & 0.1017 & 0.0948 & 0.1201\\
\hline
		\multicolumn{17}{c} {$\boldsymbol{l_X}=0.25, \quad \boldsymbol{\rho_{U_1(t)}}=0.25$}\\
		\hline
		\multicolumn{2}{c} {} &\multicolumn{5}{c} {ABias$^2$} &\multicolumn{5}{c} {AVar} &\multicolumn{5}{c} {AIMSE} \\
		\hline
		Cov$_{X_1(t)}$ &Cov$_{U_1(t)}$ & Oracle & SIMEX & RC & Average & Naive  & Oracle & SIMEX & RC& Average & Naive  & Oracle & SIMEX& RC & Average & Naive  \\
		\hline

 EXP & AR1 & 0.0005 & 0.0004 & 0.0027 & 0.0066 & 0.0688 & 0.1406 & 0.1698 & 0.135 & 0.1224 & 0.0768 & 0.1411 & 0.1703 & 0.1377 & 0.129 & 0.1456\\

EXP & CS & 0.0005 & 0.0002 & 0.0016 & 0.005 & 0.0529 & 0.1406 & 0.1634 & 0.1366 & 0.1238 & 0.0861 & 0.1411 & 0.1636 & 0.1382 & 0.1288 & 0.139\\

EXP & UN & 0.0005 & 0.0005 & 0.0019 & 0.0058 & 0.0559 & 0.1401 & 0.1666 & 0.1381 & 0.1236 & 0.0836 & 0.1406 & 0.1671 & 0.14 & 0.1294 & 0.1395\\
\hline
\end{tabular}
}
\end{table}

\begin{figure}[h]
	\centering
	\includegraphics[width=18cm]{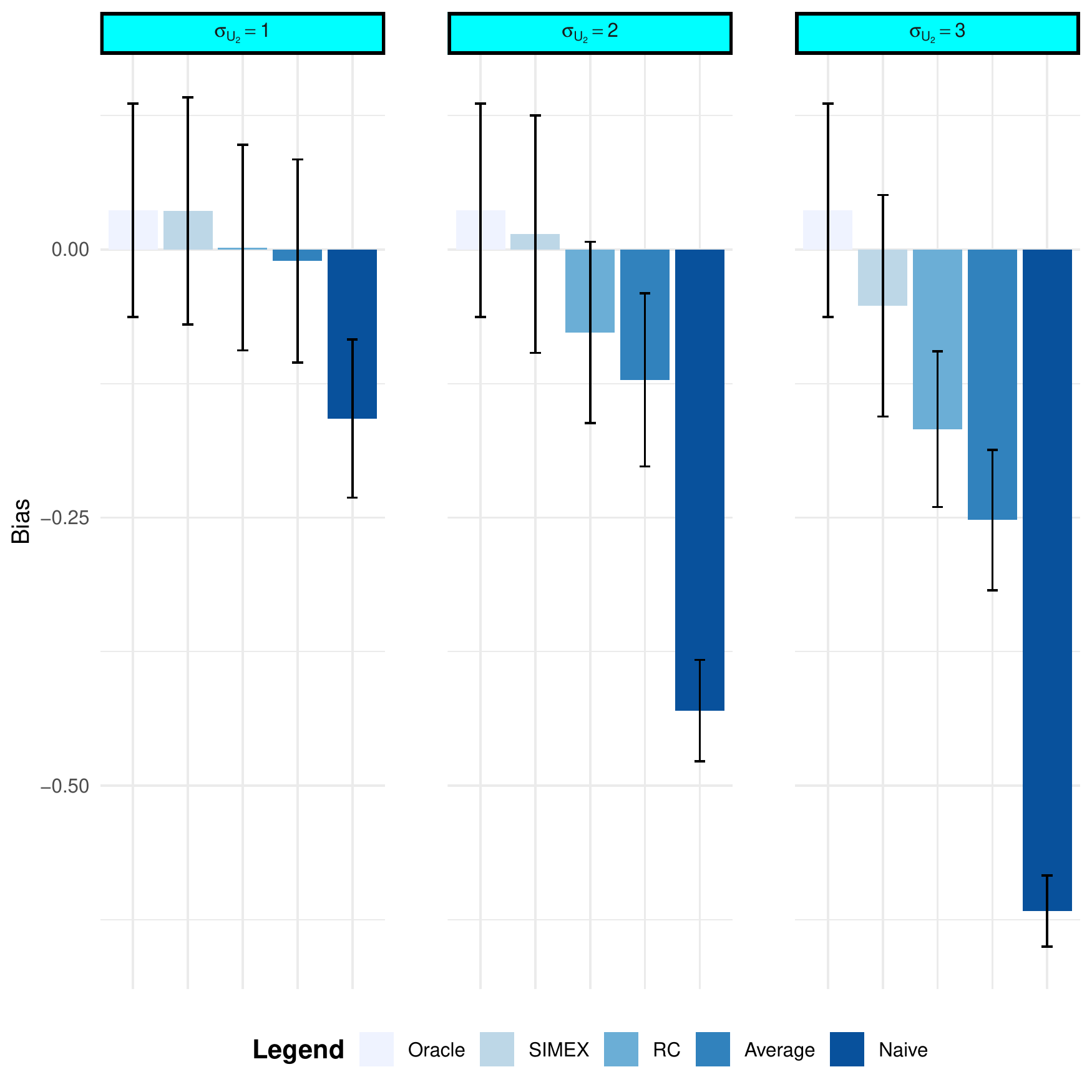}
	\caption{Bar plots of the bias of the scalar regression coefficient ($\widehat{\beta_2}$) in the generalized functional linear function from simulations of the oracle, SIMEX, RC, average, and naive estimators. The length of an error bar represents twice the standard deviation of the corresponding $\widehat{\beta_2}$. In these simulations, $n=2,000$, $\sigma_{X_1}=3$, $l_X=0.15$, $l_U=0.15$, $\sigma_{X_2}=4$, and $\sigma_{U_2}= 1,2,$ or $3$. 
	\label{barplot}}
\end{figure}

\section{Application}	

The Centers for Disease Control and Prevention established the National Health and Nutrition Examination Study (NHANES)\cite{NHANES} in the early 1960s with the goal of evaluating the health and nutritional status of Americans. NHANES collects data through interviews, questionnaires, and medical examinations, including information on demographics, socioeconomic status, dietary habits, health behaviors, medical conditions, physiological examination results, and laboratory assessments of blood samples. In the 2011-2014 cycles, NHANES collected, physical activity data, including physical activity intensity, with ActiGraph triaxial accelerometers (model GT3X+, manufactured by ActiGraph of Pensacola, FL). NHANES participants wore this device for 24 hours a day for seven consecutive days, including during water-related activities for up to 30 minutes, as this device is water-resistant for that duration\cite{NHANES}. The wearable devices detect and measure the magnitude of acceleration, or "intensity," of movement and record activity every $\frac{1}{80}$th of a second (80 Hz). NHANES aggregated the 80 Hz accelerometer measurements over every minute, hour, and day for each participant \cite{NHANES}.


NHANES participants reported their health statuses, including whether they had previously been diagnosed with type 2 diabetes (T2D), in a questionnaire. We classified participants who answered "yes" as having T2D. Such participants also reported how old they were when diagnosed. We included in our analysis only those participants who responded to the question about T2D diagnosis. In a dietary interview, participants were asked for detailed dietary intake information, which was used to estimate their total energy intake (in Kcal) over the course of two days. Finally, in a further questionnaire, participants reported their demographic characteristics, including  age, sex, and race/ethnicity\cite{NHANES}. 

  
Our final analytic sample consisted of 5,021 participants aged $20$ to $80$ years who had wearable device-based physical activity data for at least four days per week during the 2011-2014 survey cycles. The sample included $52\%$ women, and participants' average age was $49$ years (standard deviation = $18$ years). Twelve percent were black, $14\%$ were Hispanic, $66\%$ were white, and $7\%$ were of another race. 

We estimated the following functional logistic regression model with measurement error: 
$$
\begin{aligned}
\log \left(\frac{P_i}{1-P_i}\right)&=\int_{0}^{1} \beta_1(t) X_{1i}(t) dt+ \beta_2 X_{2i} +\mathbf{Z}_{i} \boldsymbol{\alpha}, \\
W_{1ij}(t) &=X_{1i}(t)+U_{1ij}(t), \text{and} \\
W_{2ij} &= X_{2i} + U_{2ij}.
\end{aligned}
$$
$P_i$ represents the probability of having T2D for the $i$th participant, and $X_{1i}(t)$ and $X_{2i}$ are the true measures of physical activity intensity and total caloric intake for participant $i$, respectively. The observed covariates, $W_{1ij}(t)$ and $W_{2ij}$, represent the observed wearable device-based measure of physical activity intensity and self-reported total caloric intake, respectively, for the $i$th subject on the $j$th day. Similarly, $U_{1ij}(t)$ and $U_{2ij}$ are the measurement errors associated with the device-based measure of physical activity intensity and total caloric intake, respectively, for the $i$th subject on the $j$th day. Finally, $\boldsymbol{Z}_i$ represents a vector of error-free covariates including age, race, and gender for the $i$th subject.

\subsection{Application results}

\begin{figure}[h]
	\centering
	\includegraphics[width=15cm]{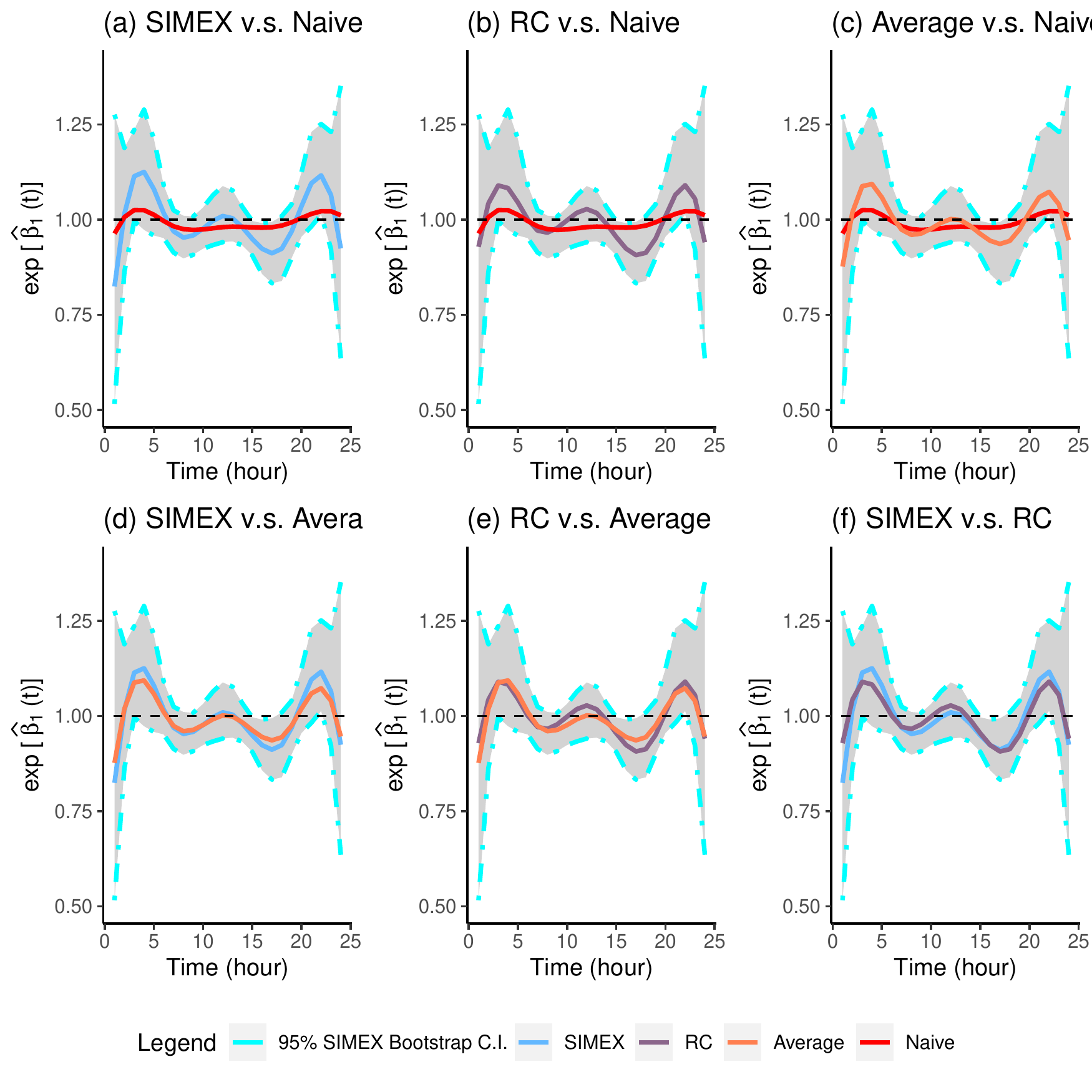}
	\caption{Plots comparing the SIMEX, RC, average, and naive estimates of the exponentiated coefficients for the relationship between physical activity intensity and T2D [$\exp\{\widehat{\beta_1}(t)\}$] in the 2011-2014 NHANES data after adjusting for total caloric intake, age, sex, and race.\label{app2011}}
\end{figure}

\begin{figure}[h]
	\centering
	\includegraphics[width=15cm]{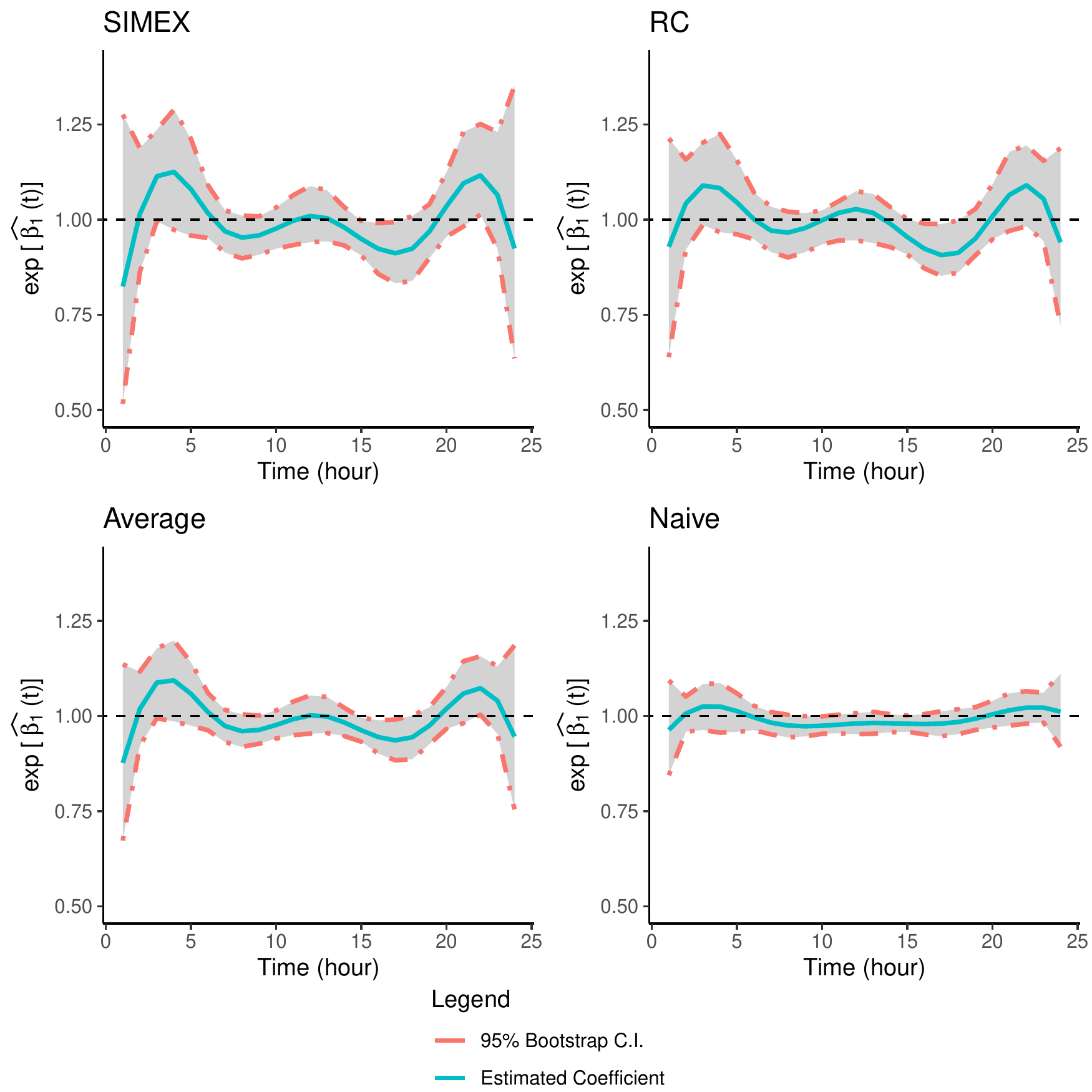}
	\caption{Plots of the $95\%$ point-wise bootstrap confidence intervals (CI) of exponentiated coefficients of the association between physical activity intensity and T2D [$\exp{\widehat{\beta_1}(t)}$] in the 2011-2014 NHANES data after adjusting for total caloric intake, age, sex, and race.\label{boot2011}}
\end{figure}

Table \ref{Table:errorfree} shows the estimated exponentiated regression coefficients and their corresponding $95\%$ confidence intervals from the SIMEX, RC, average, and naive estimators. We computed the $95\%$ confidence intervals with a nonparametric bootstrap with $500$ bootstrap samples for each of the four estimated models. The table excludes the time-varying coefficients for physical activity intensity, which we summarize graphically below (See Figure \ref{app2011} and \ref{boot2011}). Table \ref{Table:errorfree} reveals that measurement error in self-reported total caloric intake attenuated its association with T2D because the intake coefficients in the models with the naive and average estimators are closer to 1 than those in the models with the SIMEX and RC estimators. This aligns with the results of our simulations (Figure \ref{barplot}). In addition, we observed an inverse association between the total caloric intake and T2D after adjusting for physical activity, age, sex, and race/ethnicity. Moreover, Table \ref{Table:errorfree} shows that measurement error can influence the estimated associations between error-free covariates and T2D diagnosis, with the direction of such influence depending on the error-free covariate. For example, error correction with the SIMEX and RC estimators reduced the coefficient for age relative to the models with the naive and average estimators that do not correct for measurement error. In contrast, error-corrected estimators increased the magnitude of the coefficient for sex compared to the models with the error-uncorrected estimators. The difference in odds of T2D for whites and Hispanics was not statistically significant in the models with the SIMEX and RC estimators, but was significantly different in the model with the average estimator. 

\begin{figure}[h]
	\centering
	\includegraphics[width=15cm]{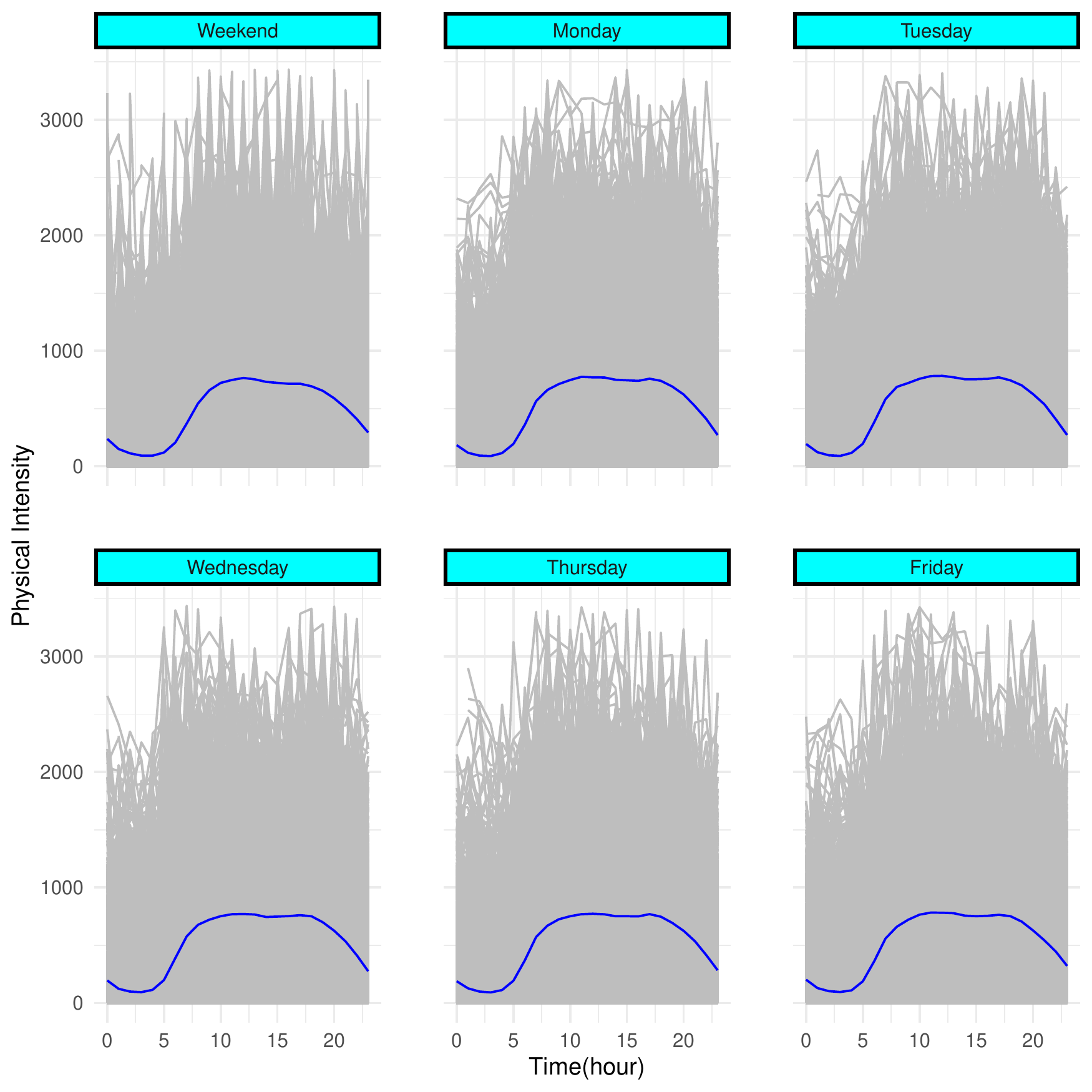}
	\caption{Wearable device-based physical activity intensity from the 2011-2014 cycles of the NHANES plotted against time (hour) for the weekend and individual weekdays. The grey lines in the background are the trend of physical intensity over time for each participant and the blue line is the average physical intensity across all participants over time for each day. \label{PA2011}}
\end{figure}

Figure \ref{PA2011} shows that the intra-day pattern of device-based physical activity intensity was similar across individual weekdays and weekends. This similarity allowed us to estimate the measurement error covariance matrix and consequently approximate conditional distributions for true physical activity intensity given the observed data with information from both weekdays and weekends. Figure \ref{app2011} shows plots of $\widehat{\beta}(t)$ over the 24-hour day for each pair of estimators. The coefficient for the association between physical activity and T2D varies by time of day. Estimates based on the naive estimator, and to a lesser extent, the average estimator, strain toward the null relative to estimates based on the error-corrected estimators, SIMEX and RC.

Figure \ref{boot2011} shows plots of the nonparametric $95\%$ point-wise bootstrap confidence intervals for each of the estimators. There is a statistically significant moderate negative association between device-based physical activity intensity and T2D between the $15$th and $17$th hours (late afternoon, prior to typical dinner time) in models based on the SIMEX, RC, and average estimators. Similarly, we observed a negative nonsignificant association between physical activity intensity and T2D between the $7$th and $10$th hours (mid-morning, before typical lunchtime). In addition, there are notable positive associations between physical activity intensity and T2D during the night hours.  

From the application to the NHANES year 2011-2014 data, we found that measurement error in error-prone covariates, including dietary intake and physical activity, tended to attenuate their associations with T2D. The impacts of measurement error on error-free covariates depends on the covariates. The SIMEX estimators of both error-prone and error-free covariates tended to be the most different from the naive estimators, indicating that the SIMEX estimators were the "better" estimators among all estimators proposed in this study. This finding also aligns with the results of our simulation study. 

\begin{table}
	\caption{Estimated exponentiated coefficicents of the associations between the scalar and error-free covariates and T2D in the 2011-2014 NHANES data, based on the SIMEX, RC, average, and naive estimators.\label{Table:errorfree}} 
	\scalebox{0.83}{
	\begin{tabular}{|cc| rr rr rr rr |} 
		\hline
		\textbf{Covariates} & & \textbf{SIMEX}  & \textbf{$95\%$ C.I.} & \textbf{RC}  & \textbf{$95\%$ C.I.} & \textbf{Average} & \textbf{$95\%$ C.I.} & \textbf{Naive} & \textbf{$95\%$ C.I.}  \\
Intercept && 0.0254 &(0.0084, 0.0711 ) &0.0252 &(0.0091, 0.0624) & 0.0140 &(0.0073, 0.0349) & 0.0158 &(0.0063, 0.0292)\\
Dietary Intake && 0.9995 &(0.9991, 0.9998 ) & 0.9995 &(0.9991, 0.9998) & 0.9997 &(0.9995 ,0.9999) & 0.9998 &(0.9996 ,0.9999)\\

Age&  & 1.0461 & (1.0379, 1.0552) &1.0474  & (1.0404 ,1.0569) &1.0488 &(1.0404 1.0569)& 1.0481 &(1.0411, 1.0579)\\

Sex&&&&&&&&& \\
& Female & 0.7488 & (0.5437, 1.0302 ) & 0.7985 & (0.6041 ,1.0856  ) &0.8023 &(0.6063, 1.0848)& 0.8081 &(0.5925, 1.0952)\\

Race &&&&&&&&&  \\
& Black & 1.7061 & (1.1231, 2.5838) & 1.7467 & (1.1849 ,2.6601) &1.8152&(1.1710, 2.6608)& 1.7676 &(1.2191, 2.6464)\\

& Hispanic & 1.5907 & (0.9847 ,2.3929) & 1.5040 & (0.9930 ,2.2991) &1.5115&(1.0090, 2.3713) & 1.5895 &(0.9631 ,2.2333)\\

&Other & 0.9094 & (0.4927, 1.5227  ) & 0.9437 & (0.5242, 1.5656) &  0.9454 &(0.5173, 1.5783)& 0.9413 &(0.5233 ,1.5932)\\
\hline
\end{tabular}}
\end{table}

\section{Discussion}

Prior approaches to correcting measurement error in functional covariates in generalized functional linear regression are based on the assumption of discrete measurement error or do not distinguish measurement errors with complex error structures from random noise in the data. We developed SIMEX and RC approaches to correcting measurement error and used a repeated measures approach as an additional identifying method. We assumed the observed functional and scalar covariates {$W_{1}(t)$ and $W_{2}$} are unbiased surrogates, prone to classical additive measurement error, for the true latent functional and scalar covariates {$X_{1}(t)$ and $X_{2}$}, respectively. Both approaches allow correlation between measurement errors at different locations along the functional continuum. The SIMEX approach does not rely on any assumptions about the distribution of the functional covariate and structure of the measurement error covariance matrix. In our simulations, the SIMEX estimator had the least bias, apart from the oracle estimator, across the range of conditions. The RC estimator had the next least bias, with the average and naive estimators performing more poorly. The bias in the SIMEX estimator increased with increasing measurement error but decreased with increasing sample size (except for scalar covariates with relatively high measurement error). With increasing sample size, the bias in the RC estimator decreased for the functional covariate but increased for the scalar covariate. Bias in the RC estimator also increased with increasing measurement error for both functional and scalar covariates. The SIMEX estimator performed especially well under varying structures of functional covariate's variance-covariance matrix, effectively mirroring the very small bias of the oracle estimator. The comparatively low bias of the SIMEX and RC estimators, however, came at the price of variance higher than that for the average and naive estimators in the conditions we examined. In addition, varying correlations and magnitudes of the functional measurement error had essentially no impact on the bias or variance of the scalar covariate for any estimator. This is consistent with the property of generalized linear regression that bias in the estimator of the regression coefficient of one predictor is not affected by the measurement error in another predictor unless they are correlated or their measurement errors are correlated \cite{buonaccorsi2010measurement}. 

We applied our methods to data on physical activity intensity, dietary intake, and demographics in relation to T2D from the 2011-2014 NHANES. Estimation with the SIMEX and RC error correction estimators reduced the associations for some covariates, but increased them for others, particularly physical activity intensity, in comparison to estimation with the average and naive estimators. The association between physical activity intensity and T2D varied by time of day. Physical activity before meal times was inversely related to T2D, while physical activity at night was positively related to T2D. However, this was not statistically significant in our analyses. The inverse patterns in the association before mealtimes may be due to the fact that physical activity assists the body to use all available insulin for glucose uptake from the blood \cite{aird2018effects,vieira2016effects}, which reduces blood glucose uptake after eating. Blood glucose is significantly lower during exercise before eating than during exercise after eating \cite{aird2018effects,vieira2016effects}. This successful application suggests our approaches to correcting measurement error may be useful for generalized functional linear regression in other empirical settings as well.

\section{SOFTWARE}
The R codes for our simulations and brief instructions on how to use them are available in the GitHub repository at https://github.com/YLuan-git/GLF-SIMEX-RC. The data that support the findings of this study are also provided as part of the GitHub repository.

\section{Acknowledgments} 
This research was supported by an award from the National Institutes of Diabetes, Digestive, and Kidney Disease Award number R01DK132385. This research was also supported in part by Lilly Endowment, Inc., through its support for the Indiana University Pervasive Technology Institute.

The authors would like to thank Dr. Devon Brewer for his constructive feedback and for assistance with editing the manuscript.

\bibliography{Disser_Ref}%
\include{bibliography}





\end{document}